\documentclass[preprints,article,accept,pdftex,moreauthors]{Definitions/mdpi} 

\firstpage{1} 
\pubvolume{1}
\issuenum{1}
\articlenumber{0}
\pubyear{2025}
\copyrightyear{2025}
\datereceived{ } 
\daterevised{ } 
\dateaccepted{ } 
\datepublished{ } 
\hreflink{https://doi.org/} 

\usepackage{graphicx}
\usepackage{epstopdf}
\newcommand{\beq}{\begin{eqnarray}}
\newcommand{\eeq}{\end{eqnarray}}
\newcommand{\beqa}{\begin{align}}
\newcommand{\eeqa}{\end{align}}

\newcommand{\Slash}[1]{{\ooalign{\hfil/\hfil\crcr$#1$}}}
\newcommand{\tr}{{\rm tr}}
\newcommand{\Tr}{{\rm Tr}}

\newcommand{\Nc}{N_{\rm c}}

\newcommand{\vp}{\vec{p}}
\newcommand{\vq}{\vec{q}}
\newcommand{\vk}{\vec{k}}

\newcommand{\la}{\langle}
\newcommand{\ra}{\rangle}

\newcommand{\calL}{\mathcal{L}}
\newcommand{\calR}{\mathcal{R}}

\newcommand{\calA}{\mathcal{A}}

\newcommand{\calC}{\mathcal{C}}
\newcommand{\calD}{\mathcal{D}}

\newcommand{\calG}{\mathcal{G}}

\newcommand{\calP}{\mathcal{P}}

\newcommand{\rmd}{\mathrm{d}}
\newcommand{\rmi}{\mathrm{i}}
\newcommand{\rme}{\mathrm{e}}

\newcommand{\rmp}{\mathrm{p}}
\newcommand{\rma}{\mathrm{a}}
\Title{Thermal Quarks and Polyakov Loops in Two-Color Dense QCD}

\TitleCitation{Thermal Quarks and Polyakov Loops in Two-Color Dense QCD}

\Author{ Yugo Kurebayashi 
 $^{1,\dagger}$, Toru Kojo $^{1,2,3,*,\dagger}$ and Daiki Suenaga $^{4,\dagger}$}

\AuthorNames{Yugo Kurebayashi, Toru Kojo and Daiki Suenaga}

\isAPAStyle{%
       \AuthorCitation{Lastname, F., Lastname, F., \& Lastname, F.}
         }{%
        \isChicagoStyle{%
        \AuthorCitation{Lastname, Firstname, Firstname Lastname, and Firstname Lastname.}
        }{
        \AuthorCitation{Kurebayashi, Y.; Kojo, T.; Suenaga, D.}
        }
}

\address{%
$^{1}$ \quad Department of Physics, Tohoku University, Sendai 980-8578, Japan
\\
$^{2}$ \quad Theory Center, IPNS, High Energy Accelerator Research Organization (KEK), 1-1 Oho, 
\linebreak Tsukuba 305-0801, 
 Japan\\
$^{3}$ \quad Graduate Institute for Advanced Studies, SOKENDAI, 
 1-1 Oho, Tsukuba 305-0801, Japan \\
$^{4}$ \quad Kobayashi-Maskawa Institute for the Origin of Particles and the Universe, Nagoya University,\linebreak Nagoya 464-8602, Japan
}

\corres{Correspondence: 
  torukojo@post.kek.jp }

\firstnote{ These authors contributed equally to this work. } 

\abstract{
We study confinement and deconfinement in dense two-color QCD by analyzing the dynamics of thermal quarks and gluons. 
The Polyakov loop is used as a probe of the relevant thermal excitations, distinguishing quark- and hadron-dominated regimes in dense matter. 
To describe the Polyakov loop, we adopt both lattice-informed phenomenological models and the massive Yang–Mills framework.
After calibrating these models at zero density, we investigate in-medium modifications of 
the Polyakov loops and gluon propagators at finite temperature and density.
Diquark gaps control the screening at zero temperature, whereas the screening due to thermal quarks is sensitive to the Polyakov loop.
Inclusion of the Polyakov loop helps to reproduce lattice data at low temperature,
suggesting that thermal excitations are predominantly hadronic rather than uncorrelated quarks.
}


\begin{document}


\section{Introduction}

Recently, QCD-like theories such as two-color QCD (QC$_2$D) and QCD at finite isospin chemical potential (QCD$_I$) 
have attracted much attention as laboratories to test methodologies and concepts concerning dense matter in quantum chromodynamics (QCD) \cite{Kogut:1999iv,Kogut:2000ek,Suenaga:2022uqn,Suenaga:2023xwa,Kawaguchi:2024iaw,Suenaga:2025sln,Sun:2007fc,Brauner:2009gu,Strodthoff:2013cua,Strodthoff:2011tz}.
For these theories, lattice Monte Carlo simulations can be carried out without encountering the sign 
problem~\cite{Itou:2025vcy,Iida:2024irv,Iida:2019rah,Iida:2020emi,Boz:2019enj,Boz:2013rca,Cotter:2012mb,Hands:2011ye,Astrakhantsev:2020tdl,Bornyakov:2020kyz,Muroya:2002ry,Abbott:2023coj,Abbott:2024vhj,Brandt:2022hwy,Brandt:2017oyy}, 
and the results have been used to assess the reliability of effective models as well as weak-coupling approaches, with~or without pairing effects~\cite{Fujimoto:2024pcd,Fujimoto:2023mvc,Fukushima:2024gmp}.
In recent years, much effort has been devoted to studies of the equation of state (EOS) of dense matter. 
Notably, several concepts relevant to neutron star EOS—such as the appearance of a sound-speed peak, 
the stiffening of matter due to pairing effects~\cite{Kojo:2014rca,Kojo:2021wax,Geissel:2025vnp,Geissel:2024nmx}, 
and the significance of quark effects prior to hadron overlap~\cite{Kojo:2021hqh,Chiba:2023ftg,Kojo:2025vcq}—have been confirmed or supported by lattice results in QCD-like~theories.

In QC$_2$D and QCD$_I$ at zero temperature,
simple models that incorporate both diquark (meson) and quark degrees of freedom—such as quark–diquark (meson) models~\cite{vonSmekal:2012vx,Kamikado:2012bt,Strodthoff:2013cua}—provide good agreement with lattice results~\cite{Andersen:2018qkq,Chiba:2023ftg,Chiba:2024cny,Brandt:2025tkg,Carlomagno:2024xmi,Lopes:2025rvn,Ayala:2024sqm}.
Although these models do not explicitly realize confinement, they are able to capture the overall features of hadronic matter at low density and quark matter at high density.
A notable advantage is that, within~a single framework, both causality and thermodynamic stability are automatically satisfied from low to high densities,
especially in the domain where hadronic matter transforms into quark matter.
By construction, such models implement the essential features of the previous crossover approaches in which hadronic and quark EOS were interpolated phenomenologically~\cite{Masuda:2012kf,Masuda:2012ed,Kojo:2014rca,Kojo:2015fua,Baym:2019iky,Kojo:2021wax,Ayriyan:2021prr,Minamikawa:2020jfj,Fraga:2013qra,Kurkela:2014vha,Komoltsev:2021jzg,Gorda:2022jvk}.
For more explicit realization of the crossover, see, e.g.,~recent developments based on the quarkyonic 
matter picture~\cite{Kojo:2021ugu,Fujimoto:2023mzy,Kojo:2025vcq,Tajima:2024qzj,Fujimoto:2024doc,Ivanytskyi:2025cnn}.

As mentioned, simple quark-hadron models
have not directly addressed the nature of confinement--deconfinement transitions; rather this most subtle and intriguing question has been detoured.
In this paper, taking dense two-color QCD as our platform, 
we investigate confinement/deconfinement in dense matter by studying the thermal properties of dense matter.
We delineate the lattice data for the Polyakov loop and gluon propagators 
and compare them with analytic results.
Our central question is whether thermal excitations are dominated by quarks or hadrons; answering to this question helps us to understand confinement/deconfinement in dense~matter.

In QC$_2$D, quarks acquire the color-singlet gap and thus generate neither Debye nor Meissner masses~\cite{Kojo:2014vja,Suenaga:2019jjv,Kojo:2021knn,Contant:2019lwf,Contant:2017gtz}.
As a consequence, infrared gluons are protected from screening and remain confining.
With increasing density, the~protected phase space gradually shrinks, but~it disappears only asymptotically—a scenario known as asymptotic confinement~\cite{Rischke:2000cn,Cohen:2015soa}.
Quark matter with confining gluons provides a concrete realization of the quarkyonic matter proposed by McLerran and Pisarski~\cite{McLerran:2007qj}.
Although qualitative arguments for asymptotic confinement are plausible, a~detailed characterization remains lacking. 
Given the significant progress in lattice simulations since the original proposal of asymptotic confinement, 
it is timely to revisit this concept in a more quantitative manner.
In this paper, we pursue this goal through a set of analyses based on effective~models.

In thermal models of QCD, the~Polyakov loop plays a central role~\cite{Polyakov:1978vu}; see Ref.~\cite{Fukushima:2017csk} for a review.
The expectation value of the Polyakov loop is related to the free energy of an infinitely heavy quark, $\langle L \rangle$$\sim$$e^{-F_Q}$.
In pure Yang–Mills (pYM) theory, the~flux string from an isolated color charge extends to infinity, implying $F_Q \to \infty$ and hence $\langle L \rangle \to 0$.
In contrast, $\langle L \rangle$ becomes finite when the medium is populated with color fields, allowing the flux string of a heavy quark to dissolve.
In QCD with light quarks, the~string can also be terminated by dynamical quark–antiquark creation, so $\langle L \rangle$ may already be finite even before the medium is saturated with color fields.
A key question is how rapidly $\langle L \rangle$ grows with temperature, which directly reflects whether the dominant thermal excitations are hadrons or~quarks.

In this paper, we examine two models for the Polyakov loop.
The first is the so-called inverse Weiss potential, derived from nonperturbative gluon and ghost \mbox{propagators~\cite{Braun:2007bx,Fukushima:2012qa,Fukushima:2013xsa,Reinosa:2014ooa,Reinosa:2014zta}.}
In pYM theory, this model successfully reproduces the critical temperature of the confinement–deconfinement transition.
Here, we investigate how its predictions are modified once quarks are coupled.
The second model is a phenomenological construction calibrated to lattice simulation data, and~it has been widely used in 
Polyakov--Nambu--Jona--Lasinio (PNJL) and Polyakov--Quark--Meson (PQM) frameworks.
Although both models describe pYM well, the~inclusion of quarks clearly differentiates them: 
the inverse Weiss potential becomes excessively modified and deviates from lattice results, whereas the phenomenological model continues to reproduce the data.
This distinction becomes even more pronounced in the regime of dense quark~matter.

We further investigate gluon propagators at various values of $\mu_q$ and $T$,
incorporating Polyakov loop effects that suppress thermal quarks.
Lattice studies of gluon propagators in two-color dense QCD have been reported in Ref.~\cite{Boz:2018crd},
while model analyses with massive gluon propagators have been carried out both at zero temperature~\cite{Suenaga:2019jjv}
and at finite temperature with quark one-loop corrections~\cite{Kojo:2021knn}.
These analytic computations represent finite-$\mu_q$ applications of the framework
successfully developed for pYM~\cite{Tissier:2011ey,Tissier:2010ts,Reinosa:2017qtf,Gracey:2019xom}
and later extended to QCD with dynamical quarks~\cite{Pelaez:2014mxa,Pelaez:2017bhh}; see Ref.~\cite{Pelaez:2021tpq} for a review.
In the present work, we extend these earlier analyses for QC$_2$D by explicitly including the Polyakov-loop suppression of thermal quarks,
and we argue that this effect is helpful to explain the lattice data at low~temperature.

The remainder of this paper is organized as follows.
In Section~\ref{sec:polyakov_loop_potential_pYM}, we discuss Polyakov loop potentials in pYM theory within both phenomenological and massive-gluon models.
In Section~\ref{sec:quark_model}, we introduce quarks into these Polyakov loop models.
In Section~\ref{sec:QC2D_muq0}, the~model is calibrated at $\mu_q=0$ against lattice results.
In Section~\ref{sec:Polyakov_loop_matter}, we analyze the behavior of Polyakov loops at finite density.
In Section~\ref{sec:gluons_one_loop_formula}, we compute the in-medium screening of gluon propagators in a Polyakov loop background.
In Section~\ref{sec:screening_masses}, we examine the density and temperature dependence of the screening masses.
In Section~\ref{sec:gluons_at_finite_momenta}, we study gluon propagators at finite momenta.
Finally, Section~\ref{sec:summary} is devoted to discussions and~a summary.

\section{Polyakov Loop Potential in Pure~Yang--Mills} \label{sec:polyakov_loop_potential_pYM}

We begin with the gluon sector in the presence of background fields $\calA_0 \equiv \calA_0^a \frac{\, \lambda_a \,}{2}$, where $\lambda_a$ are the Pauli matrices for color.
The background field $\calA_0$ is assumed to be constant in space-time, 
and we consider amplitudes along the $\lambda_3$ direction.
The mean-field Polaykov loop is then given as follows: 
\beq
\Phi 
= \frac{1}{\, 2 \,} \tr [ \calP e^{ \rmi g \beta \calA_0^3 \lambda_3/2} ]
= \cos \frac{\, \beta \calR \,}{2} \,,~~~~~~~ 
\calR \equiv g \calA_0^3 \,.
\eeq
In the high 
 temperature limit ($\beta \rightarrow 0$), 
$\Phi$ approaches unity, corresponding to the deconfined phase.
At low temperatures, $\calA_0$ is expected to scale as $\sim$$T$.
If $\beta \calR \to \pi$ is in this limit, the~system is in the confined phase ($\Phi=0$).
Thus, as~the temperature increases, the~system undergoes a transition from the confined to the deconfined~phase.

Below, we examine two typical models for the Polyakov loop potential:
a phenomenological model commonly employed in PNJL and PQM frameworks, and~the inverse Weiss potential, which can be computed from nonperturbative gluon and ghost~propagators.

\subsection{Phenomenological~Potential}

The phenomenological potentials used in PNJL or PQM models~\cite{Fukushima:2017csk}
are constructed to reproduce lattice results both with and without quarks.
For example, the~two-color potential used by Brauner, Fukushima, and~Hidaka (BFH) \cite{Brauner:2009gu} is
\beq
\Omega_{\rm BFH} = - bT \big[\, 24 \Phi^2 \rme^{-\beta a} + \ln (1-\Phi^2) \, \big] \,,
\eeq
where $b =  (0.2105\, {\rm GeV})^3$ and $a=0.8581\, {\rm GeV}$.
The potential vanishes for the confined solution $\Phi \rightarrow 0$.
The first term, proportional to $-T e^{-\beta a}$, can be interpreted as an entropic term favoring the deconfined solution.
The second term, $\ln(1-\Phi^2)$, favors the confined solution; it diverges as $\Phi \to 1$, preventing the system from reaching complete deconfinement ($\Phi = 1$).
The logarithmic form originates from strong-coupling expansions, where the Haar measure yields a logarithmic contribution, and~it is related to the presence of ghosts.
The three-color version of this model is known to successfully reproduce lattice data at $\Nc = 3$,
and this situation is also true at $\Nc=2$, as~we see in the~following.

\subsection{Inverse Weiss~Potential}

To construct the inverse Weiss potential, we compute the effective potential in the presence of the background field $\calA_0^3$.
For the potential derived from the original perturbative propagators, see Refs.~\cite{Gross:1980br,Weiss:1980rj}.
The Lagrangian for the pYM part is
\beq
\calL_{\rm pYM} = - \frac{\, 1 \,}{\, 4 \,} G_{\mu \nu}^A G^{A \mu \nu} \,,~~~~~~~ 
G_{\mu \nu}^A = \partial_\mu A_\nu^A - \partial_\nu A_\mu^A + g f_{ABC} A_\mu^B A_\nu^C \,,
\eeq
where $f_{ABC}$ is the structure constant and $g$ is the coupling constant.
To construct the one-particle-irreducible (1PI) effective potential as a function of $\calA_0 = \calA_0^3 \frac{\lambda_3}{2}$, 
we first separate the gauge field as $A_\mu = \calA_0 \delta_{\mu 0} + a_\mu$, with $a_\mu$ being quantum fields.
The $G_{\mu \nu}$ can be rewritten as
$G_{\mu \nu}^A = [\calD_\mu a_\nu]^A - [\calD_\nu a_\mu]^A + g f_{ABC} a_\mu^B a_\nu^C$,
with the background covariant derivative defined by
$\calD_\mu a_\nu \equiv  \partial_\mu a_\nu - \rmi \calR \delta_{\mu 0} \big[ \frac{\lambda_3}{2}, \frac{\lambda_C}{2} \big] a_\nu^C$.
Using the Cartan bases $ \lambda_3/2$ and $\lambda_\pm/2 = (\lambda_1\pm \rmi \lambda_2)/2 $, one can write
\beq
\calD_\mu a_\nu
= \partial_\mu a_\nu^3 \frac{\lambda_3}{2}
+ \big( \partial_\mu - \rmi \calR \delta_{\mu 0} \big) a_\nu^- \frac{\lambda_+}{2} 
+ \big( \partial_\mu + \rmi \calR \delta_{\mu 0} \big) a_\nu^+ \frac{\lambda_-}{2}  \,.
\eeq
We impose the background gauge fixing condition, $\calD^\mu a_\mu = 0$, 
which breaks the gauge invariance of $a_\mu$ but maintains the invariance for $\calA_\mu$.

To model the non-perturbative propagators in the IR, we add a mass term for gluons
to capture the qualitative trend of the Landau gauge propagators in a vacuum;
see Ref.~\cite{Pelaez:2021tpq} for a recent review and Refs.~\cite{Gribov:1977wm,Vandersickel:2012tz}, which give the physical motivation based on the Gribov--Zwanziger framework.
We can either add a term $\propto m_g^2 a_\mu^2$, which is local but not gauge invariant, or~a term which is non-local but gauge invariant,
\beq
\calL_{m_g} = - \frac{\, m_g^2 \,}{2} a_\mu \bigg( g^{\mu \nu} - \frac{\, \partial^\mu \partial^\nu \,}{\, \partial^2 \,} \bigg) a_\nu \,.
\eeq
Including all these elements and using the Faddeev--Popov trick, the~massive YM model is
\beq
\calL_{\rm mYM} = \calL_{\rm pYM} + \frac{\, (\calD^\mu a_\mu)^2 \,}{\, 2\xi \,} + \calL_{m_g} + \bar{c} \calD^\mu \calD_\mu c + \calL_{\rm int}\,,
\eeq
where $c$ and $\bar{c}$ are ghost and anti-ghost fields.
Taking $\xi \rightarrow 0$, the~gluon and ghost propagators are (the gluon propagator $D_{\mu \nu}^{AB} \propto \la (a^A_{\mu} )^* (-k) a^B_{\nu} (k) \ra$)
\beq
[D_{AB}^{\mu \nu} (k) ]_{\rm tree}
= 
- \frac{\, \delta_{AB} \,}{\, k_A^2 - m_g^2 \,} P^{\mu \nu}_{k_A} 
\,,~~~~~~~~
[D^{AB}_{c} (k) ]_{\rm tree}
= - \frac{\, \delta^{AB} \,}{\, k_{A}^2 \,} \,,
\eeq
where $\delta_{AB} = 1$ for $(A,B) = (3,3), (+,+), (-,-)$, with the other components being zero. 
The projection operator for the transverse component is $P^{\mu \nu}_k = g^{\mu \nu} - k^\mu k^\nu/k^2$.
The momentum for $(3, -, +)$-components is given by
\beq
(k_{3})_{\mu} = k_\mu\,,~~~~(k_{\pm})_{\mu}  = k_\mu \pm \calR \delta_{\mu 0} \,.
\label{eq:k_def}
\eeq
With these propagators, one can construct the one-loop effective potential.
We define
\beq
F (r;m) = T \sum_n \int_{\vk} \ln \big[ ( \omega_n + r)^2 + \vk^2 + m^2 \big] \,,
\eeq
where $\omega_n = 2n \pi T (n \in {\bf Z})$ is the 
 Matsubara frequency for bosons.
The free energy densities for gluons and ghosts can be written as 
\vspace{6pt}
\beq
\Omega_g^{\rm mYM} (T;\calR)  = \frac{1}{\, 2\,}  \sum_{r=0,\pm \calR} \big[ 3F (r;m_g) + F(r;0) \big] \,,~~~~~
\Omega_{gh}^{\rm mYM} (T;\calR) = - \sum_{r=0,\pm \calR} F (r;0) \,,
\eeq
where $T$-independent constants are neglected.
The factor $3$ in the first term accounts for the trace over the projectors $P^{\mu \nu}$, while the last term arises from the longitudinal component,
which decouples from radiative corrections and hence remains massless.
Ghosts contribute with the opposite sign to~gluons.

Summing the gluon and ghost contributions, the~one-loop potential reads
\beq
\Omega_{g +gh}^{\rm mYM} (T;\calR)  = \frac{1}{\, 2\,}  \sum_{r=0,\pm \calR} \big[ 3F (r;m_g) - F(r;0) \big] \,,
\eeq
where the sum over $r$ can be explicitly evaluated as
\beq
\sum_{r=0,\pm \calR} F (r;m) 
= T \int_{\vk} \bigg(
 \ln \big[ 1 - \rme^{ - \beta E_m } \big] 
 +
 \ln \big[ 1 -2 \rme^{ - \beta E_m } (2\Phi^2 -1) +  \rme^{ - 2\beta E_m } \big]  
 \bigg) \,.
 \label{eq:gluons_sum}
\eeq
In standard perturbation theory, ghosts cancel two gluon modes, leaving only the two transverse modes.
In the present massive theory, the~massless longitudinal mode is canceled, 
but the cancellation of one mode is incomplete due to the gluon mass.
At low temperatures, massive gluons are suppressed, and~the ghost contribution, which favors the confining solution, dominates the~potential.

\begin{figure}[H]
\begin{minipage}[b]{0.48\columnwidth}
\includegraphics[width=6.5 cm]{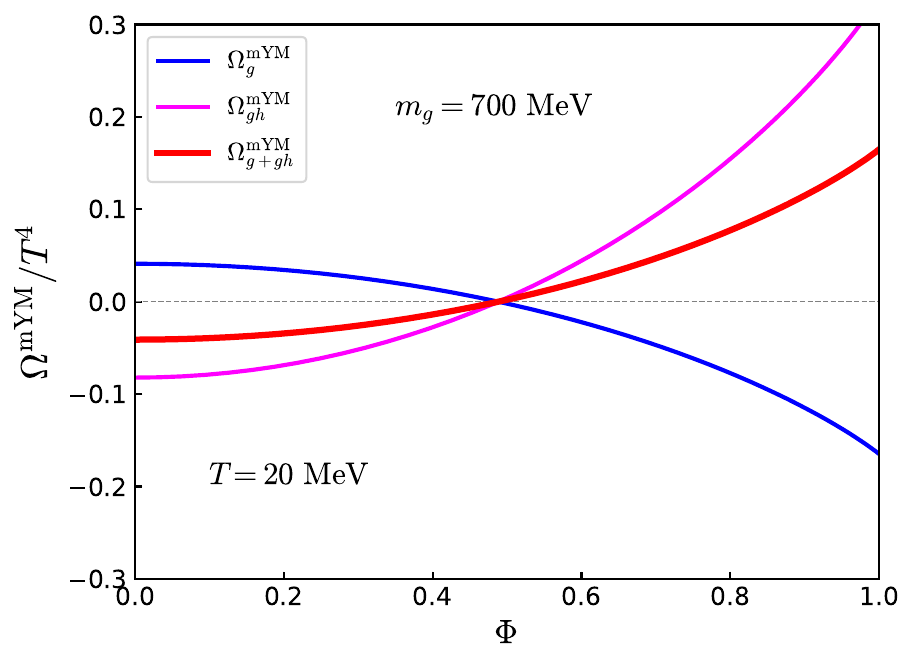}
\end{minipage}
\begin{minipage}[b]{0.48\columnwidth}
\includegraphics[width=6.5 cm]{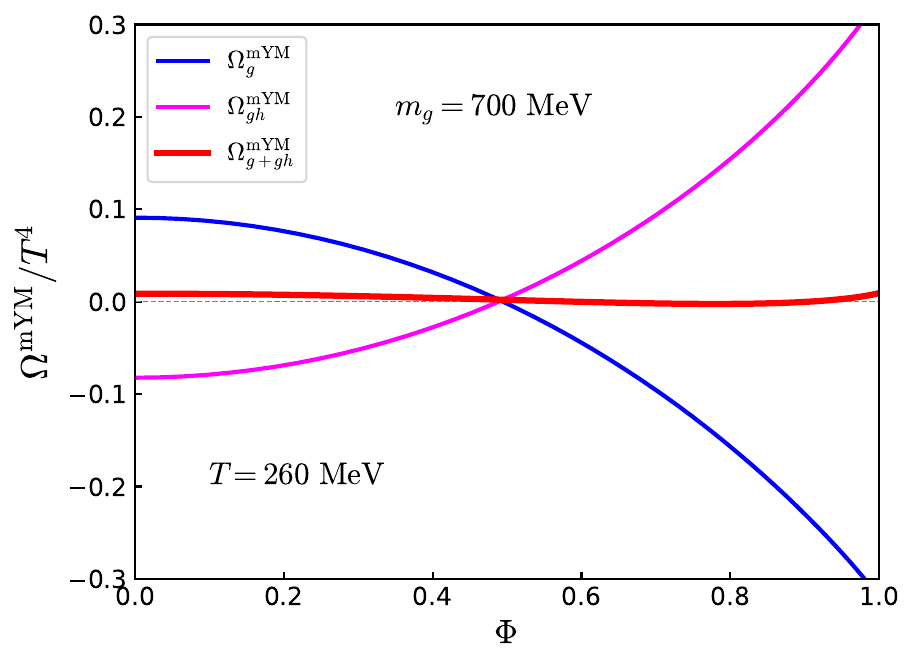}
\end{minipage}

\caption{ The Polyakov 
 loop potential before (\textbf{left}) and just after (\textbf{right}) the deconfinement transition. 
The results at $T=$ 20 and 260 MeV are shown for illustrative purposes.
The potential is normalized by $T^4$. The~normalized potential for ghosts does not depend on $T$.
The absolute value of the gluonic part grows with $T$, shifting the minimum from $\Phi=0$ to $1$.
}
\label{fig:potential_mYM_g_gh}
\end{figure}

The gluon and ghost potentials are shown in Figure~\ref{fig:potential_mYM_g_gh} for low temperature ($T=20$~MeV, confined) and high temperature ($T=260$ MeV, deconfined).
It should be emphasized that the Polyakov loop in the fundamental representation does not suppress thermal excitations in the adjoint representation.
Physically, dynamical gluons screen adjoint charges and form color-singlet states.
Indeed, the~logarithmic terms with Boltzmann factors can be rewritten as $\Tr_{\rm adj.} \big[ 1- L_{\rm adj} \rme^{-\beta E} \big]$.
Here, $L_{\rm adj}$ denotes the Polyakov loop operator in the adjoint representation, which
can be expressed in terms of the fundamental representation $L_{\rm fund}$
through a group-theoretic relation, $(N_c^2-1) L_{\rm adj} = N_c^2 L_{\rm fund} \overline{L}_{\rm fund} - 1$.
At the mean-field level, a~vanishing Polyakov loop for quarks ($\Phi=0$) necessarily implies a non-vanishing Polyakov loop for gluons and ghosts.
As a result, the~thermal pressure (=$-\Omega^{\rm mYM}$) is unphysically large at low~temperature.

In addition, we examine the temperature and $\Phi$-dependence of the mass $m_g$.
Such dependences have been studied in Ref.~\cite{Canfora:2015yia};
see also Ref.~\cite{Mei:2025kha}.
We adopt the following parametrization:
\beq
 m_g^T = \sqrt{ \big( m_g^{\rm vac} \big)^2 + c_g \Phi^{n_g} T^2 } \,,
 \label{eq:phi_dep_mg}
\eeq
which interpolates between the computations at $\Phi=0$ and $1$ within the Gribov--Zwanziger framework.
We choose $n_g=2$ and $c_g =0$ and 4.
At $\Phi=1$, the~mass at large $T$ eventually scales as $m_g$$\sim$$g_s T$,
while for $\Phi=0$, the mass is saturated, $m_g$$\sim$${\rm const}$ for $\Phi =0$.
For the vacuum mass, we choose $m_g^{\rm vac} =800$ MeV, 
which is consistent with the lattice studies for the Landau gauge propagators in pYM~\cite{Reinosa:2014ooa,Reinosa:2014zta}.

\subsection{Comparison of the Polyakov Loop~Potential}

 

\begin{figure}[H]
\begin{minipage}[b]{0.495\columnwidth}
\includegraphics[width=7.0 cm]{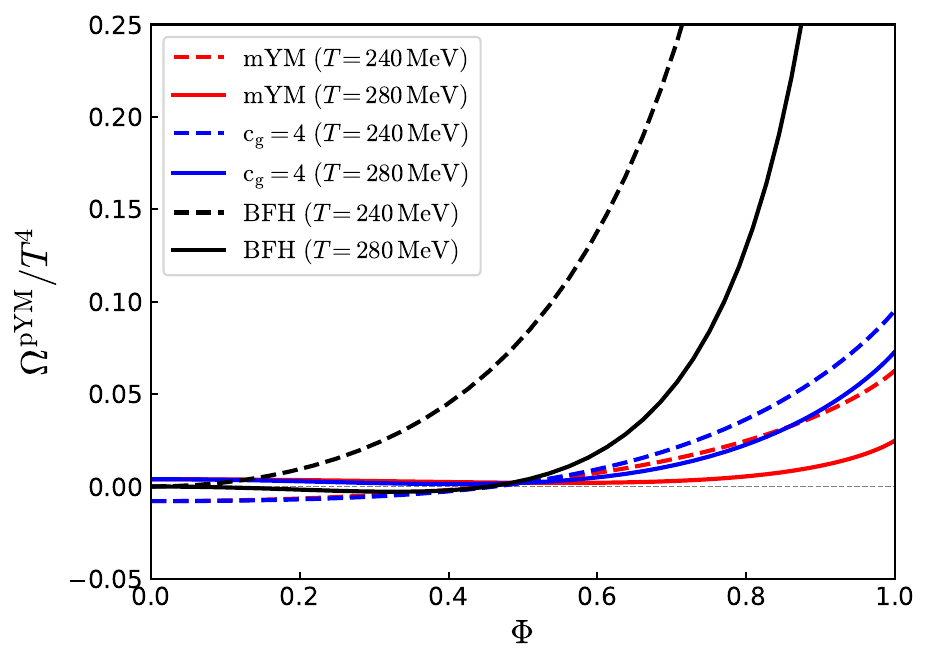}
\end{minipage}
\begin{minipage}[b]{0.495\columnwidth}
\includegraphics[width=6.7 cm]{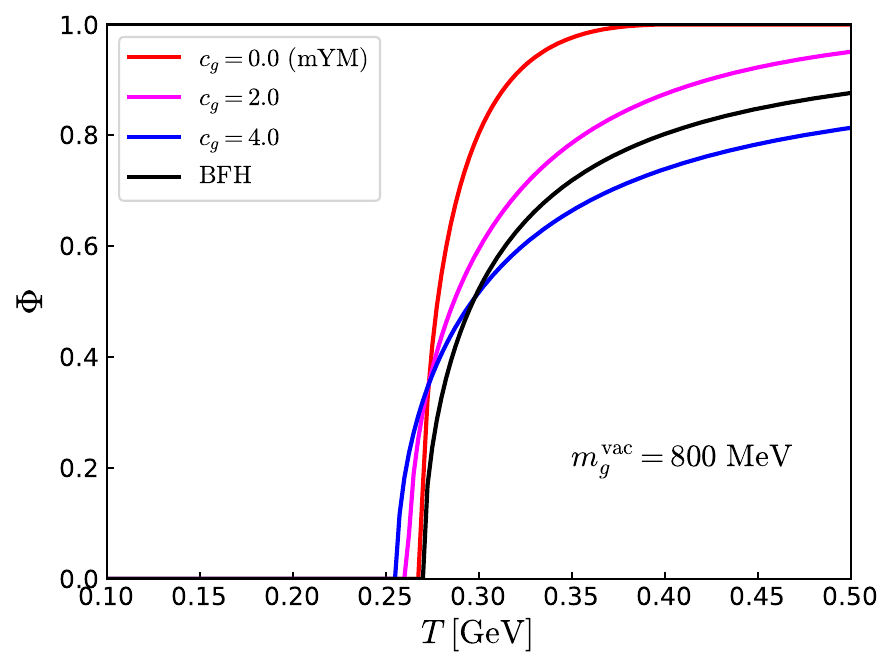}
\end{minipage}

\caption{ (\textbf{Left}) Comparisons of the Polyakov loop potential for the mYM for $m_g^{\rm vac} =800$ MeV with and without the $\Phi$-dependence of $m_g$,
and the pYM part used in the PNJL model (BFH). 
For $\Phi$-dependent $m_g^T$, we chose $c_g =4$ and $n_g =2$ in Equation~\eqref{eq:phi_dep_mg}.
All these models lead to the transition temperature being somewhere between $T=$ 240 and 280 MeV.
The potential used in the PNJL model has an energy cost around $\Phi \sim1$, much bigger than in the mYM.
 (\textbf{Right}) 
The temperature dependence of the Polyakov loop $\Phi$ for mYM models with $c_g=0,2,$ and 4, and~for the BFH model.
}
\label{fig:graph_mYM_g_gh}
\end{figure}

Here, we compare the phenomenological BFH potential with the mYM potential.
The left panel of Figure~\ref{fig:graph_mYM_g_gh} shows both potentials as functions of $\Phi$.
In both cases, the~critical temperature lies in the range 240–280 MeV.
A major difference is that the mYM potential is significantly flatter than the BFH one.
Consequently, the~Polyakov loop grows much more rapidly with temperature in the mYM case,
as illustrated in the right panel of Figure~\ref{fig:graph_mYM_g_gh}.
This rapid growth can be moderated by introducing a $T$- and $\Phi$-dependent gluon mass,
which suppresses the gluonic entropic contribution.
With $c_g \simeq 2.0$–4.0 and $n_g=2$, the~thermal evolution of the Polyakov loop in the mYM model
can be brought into qualitative agreement with the BFH result,
although the potential structure at $\Phi \gtrsim 0.5$ still shows substantial differences.
As we will see in the next section, the~behavior of the potential for $\Phi \gtrsim 0.5$~plays a crucial role once quarks are~included.

\section{Including Quarks: PNJL~Model}  \label{sec:quark_model}

\subsection{Quasi-Quark~Excitations}

We now include quarks within a quasi-particle description.
To be concrete, we use the PNJL model.
As a first step, we examine the $\mu_q=0$ case to calibrate our models against lattice data.
In particular, the~lattice data were often obtained for a nonzero diquark source
with which quarks form color-singlet diquark condensates and acquire a gap even at $\mu_q=0$. 
Without the diquark source term, the~diquark condensate is formed
only for $\mu_B \ge m_{\pi}$ (equivalently $\mu_q \ge m_\pi/2$),
as predicted by the chiral perturbation theory~\cite{Son:2000xc,Son:2000by}.



For later convenience, from~the very beginning, we derive the general expression of quark spectra at nonzero $\mu_q$ and $\Delta$.
In the presence of a temporal background field, the~kinetic term reads
\beq
\calL_{kin, q} 
= \bar{q} \big[ \rmi \Slash{\partial} + g \Slash{a}  + \calR T_3 \gamma_0 + \rmi \mu \gamma_0 - M_q \big] q
- q^T {\bf \Delta} q 
 \,,
\eeq
with $T_3 = \lambda_3/2$ and ${\bf \Delta} = \gamma_5 \tau_2 \lambda_2 \Delta$, where $\tau_2$ is the Pauli matrix in flavor space.
Employing the Nambu--Gor'kov formalism, we define $\psi = q$ and its charge-conjugated field as\linebreak $\psi_C (p) = \lambda_2 \tau_2 C \bar{q}^T (-p)$.
Then,
\beq
 \bar{\psi} (p)  \Slash{p} \psi (p) =  \bar{\psi}_C (p) \Slash{p} \psi_C (p) \,,~~~~~
 \bar{\psi} (p) \gamma_0 \calA^0 T_3 \psi (p) = \bar{\psi}_C (p) \gamma_0 \calA^0 T_3 \psi_C (p) \,,
\eeq
where we used $\lambda_2 (-T_a^T) \lambda_2 = T_a$.
The momentum space representation of the Lagrangian now reads
\beq
\tilde{\calL}_{kin, q}
=
\bar{\Psi} (p) 
\begin{bmatrix}
~\Slash{p} + \big(\calR T_3 + \rmi \mu \big)\gamma_0 - M_q  ~&~ \Delta  \gamma_5  ~\\
~ -\Delta^*  \gamma_5  ~&~ \Slash{p} +\big( \calR T_3 - \rmi \mu  \big) \gamma_0 - M_q~
\end{bmatrix}
\Psi(p) \,,
\eeq
where $\Psi(p) = (\psi (p), \psi_C (p) )^T/\sqrt{2}$.

\begin{figure}[H]

\includegraphics[width=8.0 cm]{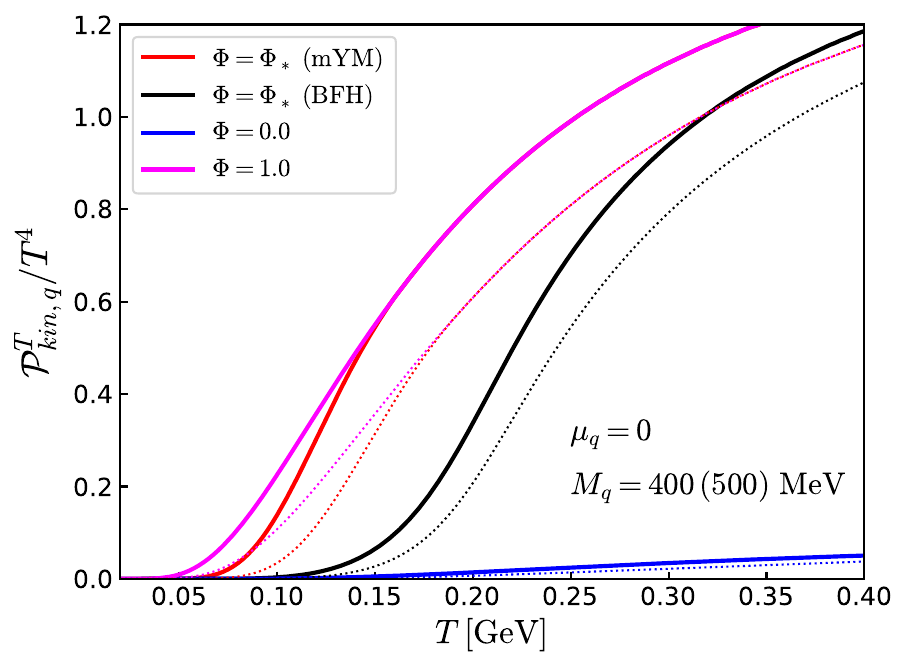}
\caption{The Polyakov loop dependence of the quark pressure in Equation~\eqref{eq:F_q^T} at $\mu_q=0$.
Here, we fix $M_q=$ 400 (500) MeV for solid (dotted) curves.
The Polyakov loop values $\Phi_*$ are dynamically determined for the mYM and BFH with quarks.
The results for fixed $\Phi=0$ and $\Phi=1$ (common for the mYM and BFH models) are also shown.
}
\label{fig:EOS_mYM_q}
\end{figure}

We note that $p_0$ and $\calR T_3$ always appear in the combination $p_0 + \calR T_3$,
while the chemical potential $\mu$ enters the quasiparticle poles as
$\xi_{\rm p,a} (\vp) = \sqrt{ (E_q (\vp) \mp \mu)^2 + \Delta^2 }$.
In the thermodynamic potential, the~shift $p_0 \rightarrow p_0 + \calR T_3$ modifies
the standard entropic logarithm according to
\beq
\tr_c \ln \big[ 1 + \rme^{-\beta (\xi_k^i +\rmi \calR T_3) } \big]
= \ln \big[ 1 + \rme^{-\beta (\xi_k^i +\rmi \calR/2) } \big] + \ln \big[ 1 + \rme^{-\beta (\xi_k^i -\rmi \calR/2) } \big] \,,
\eeq
where the trace is taken over color indices.
Combining these logarithms, the~thermal quark contribution to the free energy (or equivalently, minus the pressure) becomes as follows:
\beq
\Omega_{kin, q}^T = - \calP^T_{kin, q} 
= - 4 T \sum_{i= {\rm p,a} } \int_{\vk}  \ln \big( 1 + 2\Phi \rme^{-\beta \xi_k^i} + \rme^{-2\beta \xi_k^i} \big) 
\,.
\label{eq:F_q^T}
\eeq
The pressure at $\Phi=0$ (confined) or $\Phi =1$ (deconfined) are
\beq
\calP_{kin, q}^T 
\rightarrow \left\{
\begin{matrix}
&  4 T \sum_{ i= {\rm p,a} }
\int_{\vk}  \ln \big( 1+ \rme^{- \beta \Nc \xi_k^i} \big) &~~ (\Phi \rightarrow 0)
\nonumber \\
&  4 \Nc T \sum_{ i= {\rm p,a} }
\int_{\vk}  \ln \big( 1+ \rme^{-\beta \xi_k^i} \big) &~~ (\Phi \rightarrow 1)
\end{matrix}
\right.
\eeq
where we write $\Nc =2$.
The thermal quark pressure is  larger at least by a factor $\Nc$ in the deconfined phase than in the confined phase.
The difference would be even greater in the high temperature limit; 
for $\Phi=0$ with $\xi_k$$\sim$$|\vk|$, we rescale the variable as $\Nc |\vk| = |\vk'|$ \mbox{to get}
\beq
\calP_{kin, q}^{T\rightarrow \infty} (\Phi=0) \sim \Nc^{-4} \calP_{kin, q}^{T\rightarrow \infty} (\Phi=1) \,.
\label{eq:P_q^T_parametric}
\eeq
In order to examine the impact of the Polyakov loop, in~Figure~\ref{fig:EOS_mYM_q},
we fix $M_q=400$ MeV (later, we consider the dynamical evolution of $M_q$) and 
show the thermal quark pressure at $\mu_q=0$ with $\Phi_*$ minimizing the potentials for mYM + quarks and BFH.
We also show the $\Phi=0$ and $1$ cases for comparison.  
It is clear that the suppression effects due to the Polyakov loop are substantial;
$\calP_{kin, q}^T$ at $\Phi=0$ is much smaller than the $\Phi=1$ case, 
as anticipated from Equation~\eqref{eq:P_q^T_parametric} for the quark sector.
For dynamical computations allowing the variance of $\Phi$,
the growth of $\Phi$ is driven by the thermal quark (hadronic) excitations.
Thermal hadronic contributions rise rapidly around $T$$\sim$150–200 MeV as in the Hagedorn gas,
as also does the Polyakov loop. 
In this regime, hadrons strongly overlap, and there are color fields to which quarks couple become available to form a quark--gluon--plasma; 
here, hadrons dissociate into a quark–gluon plasma,
and quark descriptions should be~valid.

\subsection{Evolution of~Condensates}

To determine the evolution of the in-medium gaps,
we include both the zero-point energy and the potential terms associated with the gaps and the Polyakov loop.
For this purpose, we adopt the pNJL-type thermodynamic potential
\begin{align}
\Omega_q 
= - 4 \sum_{ i= {\rm p,a} }
\int_{\vk} \big[~ \xi_k^i + T \ln \big( 1 + 2\Phi \rme^{-\beta \xi_k^i} + \rme^{-2\beta \xi_k^i} \big) ~\big ] 
+ \frac{\, ( M_q -m_q )^2 + ( \Delta - \Delta_{\rm ext} )^2  \,}{\, 2 G \,} 
\,,
\end{align}
where $G$ denotes the NJL coupling constant, $m_q$ the current quark mass, and~$\Delta_{\rm ext}$ an external diquark source term.
The current quark mass $m_q$ is adjusted to reproduce the pion mass employed in the lattice simulations.
In the lattice setup, the~diquark source term $\Delta_{\rm ext}$ is introduced to stabilize the numerical calculations;
we take its value accordingly to ensure consistency with the lattice~data.

\textls[-10]{Including the pYM term, the~gap equations for $\Phi$, $M_q$, and~$\Delta$ are
($\Omega_{\rm QC_2D} = \Omega_{\rm pYM} + \Omega_q$)}
\begin{align}
\frac{\, \partial \Omega_{\rm QC_2D} \,}{\, \partial \Phi \,} 
&= \frac{\, \partial \Omega_{ {\rm pYM} } \,}{\, \partial \Phi \,} 
- 4 T \sum_{ i= {\rm p,a} }
\int_{\vk} \bigg[
 \frac{\, \Theta ( \Lambda_{\rm NJL} - |\vk| ) \,}{\,  \cosh(\beta \xi_k^i ) + \Phi  \,} 
 +  \frac{\, \Theta ( |\vk| - \Lambda_{\rm NJL}) \,}{\,  \cosh(\beta \xi_k^i \big|_{\eta=\eta_{\rm ext}} ) + \Phi  \,} 
 \bigg] 
\,, 
\nonumber \\
\frac{\, \partial \Omega_{\rm QC_2D} \,}{\, \partial \eta \,}  
&=
- 4 \sum_{ i= {\rm p,a} }
\int_{\vk} \Theta ( \Lambda_{\rm NJL} - |\vk| )
\frac{\, \partial \xi_k^i \,}{\, \partial \eta \,}
 \frac{\, \sinh (\beta \xi_k^i) \,}{\, \cosh (\beta \xi_k^i) + \Phi  \,} 
+ \frac{\, \eta - \eta_{\rm ext} \,}{\, 2G \,}
 \,,
 \label{eq:gap_eq}
\end{align}
where $\eta = M_q$ or $\Delta$.
In the low temperature limit ($\beta \rightarrow +\infty$), 
the thermal factors in the integrand depend on $\Phi$ very weakly;
 the thermal factor behaves as
\mbox{$ 1/(\cosh (\beta \xi_k^i) + \Phi) \simeq \rme^{-\beta \xi_k^i}$} for the $\Phi$ part
and $ \sinh (\beta \xi_k^i) /(\cosh (\beta \xi_k^i) + \Phi) \simeq 1$ for the $M_q$ and $\Delta$ parts.

Here, we have assumed that the gaps damp at high energy and parametrize such damping scale as $\Lambda_{\rm NJL}$.
This scale appears automatically if we use momentum-dependent interactions, e.g.,~the $\sim$$1/p^2$ type.
In the NJL model, the contact interactions yield only momentum-independent gaps.
Thus, we have to impose such a scale $\Lambda_{\rm NJL}$ by hand so that $\partial \xi_k^i/\partial \eta$ is nonzero only for $|\vk| \le \Lambda_{\rm NJL}$.
This results in the standard NJL gap equations.
Meanwhile, the~integral in the equation for $\Phi$ should include both thermal quarks with and without the effective masses,
so we should keep the integral for $|\vk| \ge \Lambda_{\rm NJL}$.

\subsection{The Pion~Mass}

To calibrate our model parameters to the lattice setup,
we first compute the pion mass within the model.
In the framework of the RPA, the~pion mass at $\mu_q = T = 0$ is \mbox{determined by}
\beq
1 - 4 G \Nc \delta_{ab} \int_{\vp}  \frac{\,  \Theta ( \Lambda_{\rm NJL} - |\vk| ) \,}{\, E_p - m_\pi^2/4E_p \,} = 0 \,.
\eeq
In the chiral limit, this equation reduces to the gap equation for $M_q$ at $m_\pi = 0$,
in accordance with the Nambu–Goldstone~theorem.

\section{The Polyakov Loop in QC$_2$D: Calibration at \boldmath{$\mu_q=0$}} \label{sec:QC2D_muq0}



We first briefly summarize the qualitative differences between the mYM and phenomenological BFH potentials after including quarks,
then we proceed to quantitative comparisons with lattice~results.

\begin{figure}[H]

\begin{minipage}[b]{0.495\columnwidth}
\includegraphics[width=7.0 cm]{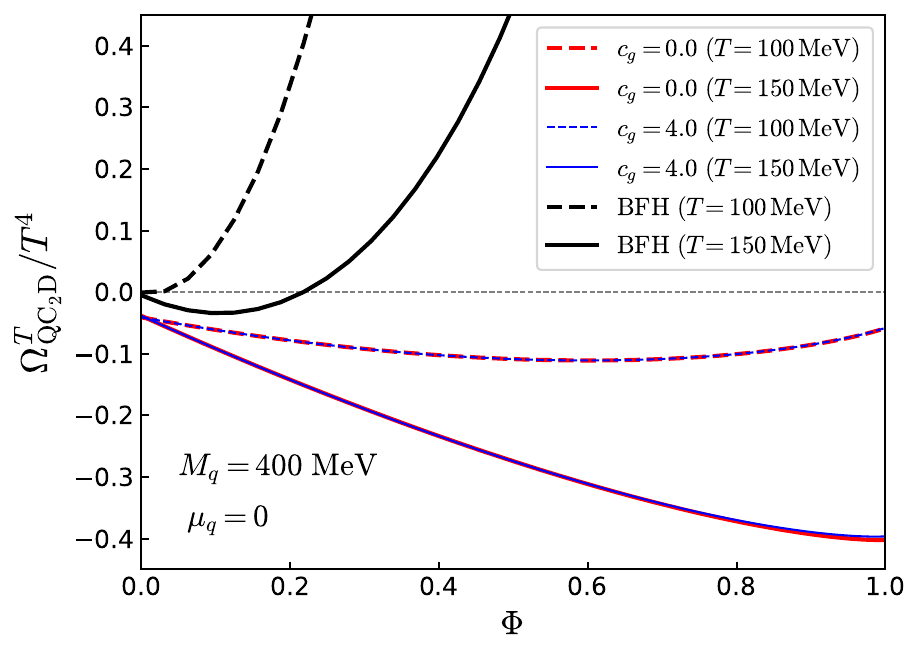}
\end{minipage}
\begin{minipage}[b]{0.495\columnwidth}
\includegraphics[width=6.8 cm]{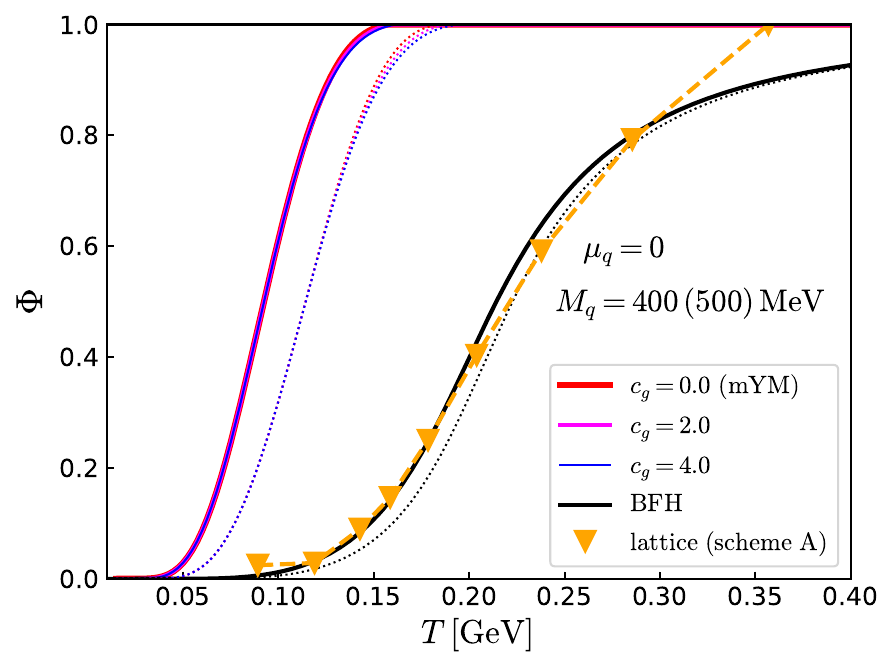}
\end{minipage}

\caption{(\textbf{Left}) The Polyakov loop potential at $\mu_q=0$ in the same setup as in Figure~\ref{fig:graph_mYM_g_gh} except that thermal quark contributions in Equation~\eqref{eq:F_q^T} are added and the temperatures are now $T=100$ and 150 MeV. The~effective quark mass is fixed to $M_q=400$ MeV.
 (\textbf{Right}) 
The temperature dependence of the Polyakov loop $\Phi$. The~solid (dotted) lines are results for $M_q=400$ (500) MeV.
The lattice results (renormalization scheme A in Ref.~\cite{Boz:2019enj}) are also plotted for comparison. 
}
\label{fig:graph_mYM_q}
\end{figure}

We begin by examining the structure of the Polyakov loop potentials at $\mu_q = 0$.
Figure~\ref{fig:graph_mYM_q} shows the Polyakov loop potential,
$\Omega^T_{\rm QC_2D} \equiv \Omega_{\rm pYM} + \Omega_q^T$, with~quarks included (left),
and the corresponding temperature dependence of the Polyakov loop (right).
Here, $\Omega_{\rm pYM}$ is replaced by either $\Omega_{gl+gh}$ for the mYM or $\Omega_{\rm BFH}$ for the phenomenological model.
For the mYM potential, we consider $c_g = 0.0$ and $4.0$ with $n_g = 2$ and $m_g^{\rm vac} = 800$ MeV.

Adding quarks to the pYM potential significantly alters its qualitative structure,
leading to distinct behaviors between the mYM and BFH models.
While both models exhibit similar transition temperatures in the pYM case,
the mYM potential is much flatter with respect to $\Phi$, so the deconfined solution is strongly favored once quarks are included.
In contrast, the~BFH potential incurs a large energy cost near $\Phi \simeq 1$ (diverging as $\Phi \rightarrow 1$) due to the Haar measure term $\sim$$\ln(1-\Phi^2)$,
which delays the growth of $\Phi$.
These differences are directly reflected in the temperature dependence of the Polyakov~loop.

We now turn to more quantitative discussions.
To compare our model calculations with lattice simulation results,
we must use model parameters adjusted to the setup used in lattice computations.
First, we note that there is no unique way to set the scale in QC$_2$D~\cite{Iida:2020emi}
since there are no real-world counterparts. 
In this work, we consider the lattice results in which the string tension is fixed as $\sqrt{\sigma_s} \simeq 440$ MeV.
For the current quark masses, the~``fine ensemble'' lattice simulations~\cite{Boz:2019enj} use a relatively heavy quark mass 
that yields a pion-to-rho mass ratio of $m_\pi/m_\rho$$\sim$$0.8$,
which is substantially larger than the corresponding ratio in the real world, $m_\pi/m_\rho \simeq 0.18$.
The heavy pion mass of $m_\pi \simeq 637$ MeV was used to reduce the computational cost for extensions to finite $\mu_q$.
We take into account this heavy current quark mass in our model~predictions.
We test two parameter sets, set I and set II, in Table \ref{table:NJL_para}.

\begin{table}[H] 
\caption{Two parameter 
 sets are used for the quark model.
We tune the current quark mass $m_q$ for a given coupling $G$ so that the RPA yields a pion pole at $m_\pi = 637$ MeV.
The cutoff $\Lambda_{\rm NJL}$ is chosen to be the same for both parameter sets.
}
\begin{tabularx}{\textwidth}{CCCCCC}
\toprule
\textbf{Name}	& \boldmath{$m_{\pi}\,[{\rm MeV}] $} & \boldmath{$\Lambda_{\rm NJL}\,[{\rm MeV}] $}  & \boldmath{$G\,[{\rm GeV^{-2}}]$}	& \boldmath{$m_q \,[{\rm MeV}] $}   &   \boldmath{$M_q \,[{\rm MeV}] $}  \\
\midrule
set I	 & 637 & 657 & 5.42	& 121 & 378 \\
set II	 &         &       & 	7.23 & 110 & 512 \\
\bottomrule
\end{tabularx}
\label{table:NJL_para}
\end{table}
\unskip

\begin{figure}[H]
\includegraphics[width=7.4 cm]{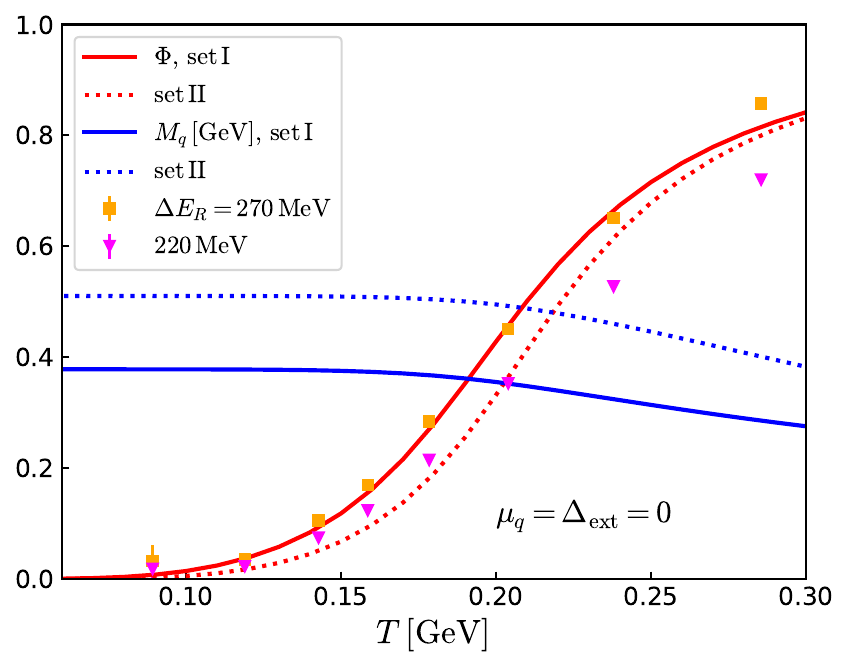}
\caption{The temperature 
 dependence of the Polyakov loop and the effective quark mass $M_q$ for the set I (solid) and II (dotted) at $\mu_q = 0$ with  $\Delta_{\rm ext} = 0$.
For the lattice data with $m_\pi \simeq 637$ MeV, we use two renormalization schemes with $\Delta E_R = E_R - E_B = 220$ and $270$ MeV
to convert the lattice data in scheme B; see Equation~\eqref{eq:reno_L_B_to_A}.
}
\label{fig:L_mu0}
\end{figure}   

Another important step is to set the renormalization factor of the Polyakov loop.
We are not aware of how to rigorously match the Polyakov loop in our model and lattice data.
For this reason, we simply choose the renormalization factor for the lattice results 
to make model results and lattice data consistent at $\mu_q=0$ (Fig.~\ref{fig:L_mu0}),~then we make model predictions for the Polyakov loop at $\mu_q \neq 0$.
Later, these model predictions are compared with the lattice~results.

The renormalized and bare Polyakov loops are related by
(in Ref.~\cite{Boz:2019enj}, the~notation ``$L$'' is used in place of $\Phi$) 
\beq
\Phi_R (T) = Z_\Phi^{N_\tau} \Phi_0 \big(  (N_\tau a)^{-1} \big)
\eeq
where $T= (N_\tau a)^{-1}$ and $Z_\Phi$ is the renormalization factor. Ref.~\cite{Boz:2019enj} discussed two renormalization schemes,
\beq
\Phi_R^{\rm scheme~A} (T = 1/4a) = 1.0 \,,~~~~~~~~
\Phi_R^{\rm scheme~B} (T = 1/4a) = 0.5 \,,
\eeq
where $1/4a \simeq 356.875$ MeV.
In Ref.~\cite{Boz:2019enj}, most results are presented in scheme B.
In the continuum expression, the~renormalization factors correspond to the choice of the zero-point energy,
\beq
Z_\Phi^{N_\tau} 
= \exp\bigg[ \beta \frac{\, \ln Z_\Phi \,}{\, a~\,} \bigg]
\equiv \rme^{ \beta (E_R - E_0 )} \,.
\eeq
The Polyakov loops in schemes A and B can be written as follows:
\beq
\Phi_R^{\rm scheme~A/B}  (T; E_R) = \rme^{ \beta (E_R^{A/B} - E_0 )} \Phi_0 (T) \,.
\eeq
In Ref.~\cite{Boz:2019enj}, most of the data are given in scheme B.
Hence, it is convenient to write the Polyakov loops at the energy scale $E_R$ as follows:
\beq
\Phi_R (T; E_R) 
= \rme^{ \beta (E_R - E_B )} \Phi_R^{\rm scheme~B} (T) \,.
\label{eq:reno_L_B_to_A}
\eeq
For instance, to~convert the data from scheme B to A, we substitute $E_R^A$ such that $E_R^A - E_R^B \simeq 247$ MeV.
For fitting to our model at $\mu_q=0$, we use $E_R \simeq 270$ MeV unless otherwise~stated.

It is important to note that these renormalization factors affect not only the magnitude of the Polyakov loop but also its temperature derivatives.
For instance, the~location of the susceptibility peak depends on the renormalization scheme.
This scheme dependence can be eliminated by considering the ratio $\Phi_R(T;\mu_q)/\Phi_R(T;\mu_q=0)$,
allowing a physically meaningful discussion of $\mu_q$ effects.
At low temperatures, this ratio is inconvenient since both numerators and denominators become vanishing.
For this reason, we do not show the ratio but plot $\Phi$ itself after calibrating $E_R - E_B$ 
so that our $\Phi$ at $\mu_q=0$ matches the lattice results.
We then examine the $\mu_q$-dependence of the Polyakov~loop.

\section{The Polyakov Loop in Baryonic and Quark~Matter} \label{sec:Polyakov_loop_matter}

\subsection*{Outline} 
\begin{figure}[H]
\includegraphics[width=8 cm]{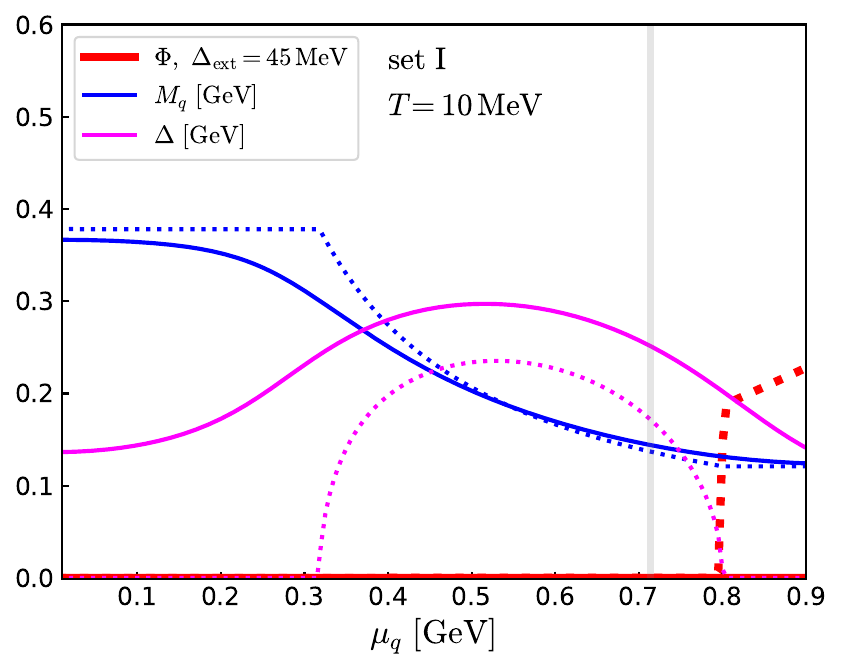}
\caption{ The $\mu_q$-dependence of the Polyakov loop, the~effective quark mass $M_q$, 
and the diquark gap $\Delta$ for the set I at $T = 10$ MeV with  $\Delta_{\rm ext} = 45$ MeV (solid) and $0$ MeV (dotted).
The reduction in $\Delta$ at large $\mu_q$ is the UV cutoff artifact,
so we compare our model results with the lattice data at $\mu_q$ less than $714$ MeV (shown as the gray line). 
For $\Delta_{\rm ext} = 45$ MeV, the~$\Delta$ is nonzero even at $\mu_q =0$,
and transitions are crossover everywhere, as the chiral and $U(1)_B$ symmetries are explicitly broken.
The nonzero Polyakov loop can be found only for $\mu_q \gtrsim 800$ MeV in the $\Delta_{\rm ext}=0$ case 
where quarks become gapless because of the UV cutoff artifact.
}
\label{fig:gap_T0010}
\end{figure}

We now explore the finite $\mu_q$ domain, beginning with the overall behavior of various condensates.
It is well established that for $\Delta_{\rm ext}=0$, a~diquark condensate forms when $\mu_q \ge m_\pi/2$.
Initially, the~diquarks are of the Bose--Einstein condensation (BEC) type, forming molecule-like pairs,
but with increasing density, they evolve into Bardeen--Cooper--Schrieffer (BCS)-type pairs near the quark Fermi surface.
For finite $\Delta_{\rm ext} \neq 0$, diquark condensates can form even before $\mu_q = m_\pi/2$.
Figure~\ref{fig:gap_T0010} shows $M_q$, $\Delta$, and~$\Phi$ at a fixed temperature $T=10$ MeV for various $\mu_q$ using parameter set I.
The solid (dotted) lines correspond to $\Delta_{\rm ext} = 45$ MeV, which matches $ja=0.03$ in the lattice results of Ref.~\cite{Boz:2019enj}.
As mentioned, nonzero $\Delta_{\rm ext}$ allows the diquark condensate to form even for $\mu_q < m_\pi/2$.
Meanwhile, the~reduction of $\Delta$ at very large $\mu_q$ is simply an artifact of the UV cutoff;
beyond the Fermi surface, with~increasing $\mu_q$, the available phase space for pairing diminishes because of the UV cutoff;~hence, diquarks cannot form at very large densities.
The gap $\Delta$ vanishes around $\mu_q \simeq 800$ MeV, leaving quarks gapless on the Fermi surface.
In this regime, even a small temperature of $T$$\sim$$10$ MeV is sufficient to induce substantial Polyakov loop expectation~values.

\begin{figure}[H]
\includegraphics[width=9 cm]{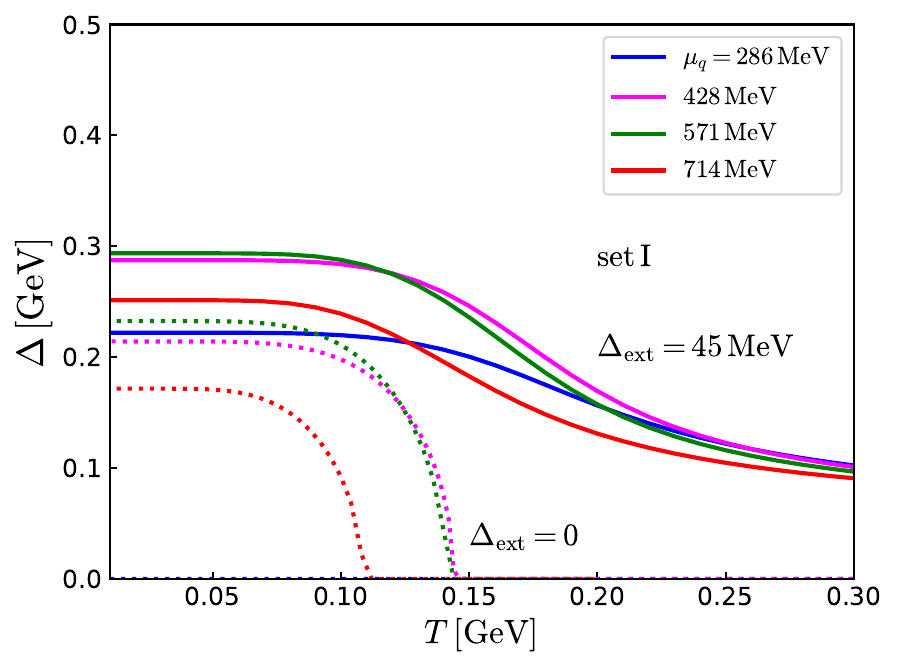}
\caption{ The temperature dependence of the diquark gap $\Delta$ for set I with  $\Delta_{\rm ext} = 45$ MeV (solid) and $0$ MeV (dotted).
}
\label{fig:Del_Tvary}
\end{figure}

Figure~\ref{fig:Del_Tvary} shows the diquark gaps $\Delta$ for $\mu_q = 286, 428, 571,$ and $714$ MeV as a function of temperature.
For $\Delta_{\rm ext} = 0$ MeV, parameter set I predicts a second-order transition at $T_s$$\sim$$110$–140 MeV.
In Ref.~\cite{Boz:2019enj}, the results at finite $j$ are extrapolated to $j\rightarrow 0$, giving an estimate $T_s \simeq 100$–120 MeV, which is roughly consistent with set I.
In contrast, set II, used with the BFH potential, predicts $T_s \simeq 170$–200 MeV, indicating a different definition of the theory from that in Ref.~\cite{Boz:2019enj}.
This explains why the NJL coupling $G$ in set I is approximately 0.75~times smaller than in the BFH~setup.
For comparisons, in figure 8 we also show the results for fixed values of Polyakov loop at $\Phi=0$ and $1$.

\vspace{-6pt}
\vspace{-6pt}
\begin{figure}[H]
\includegraphics[width=6.9 cm]{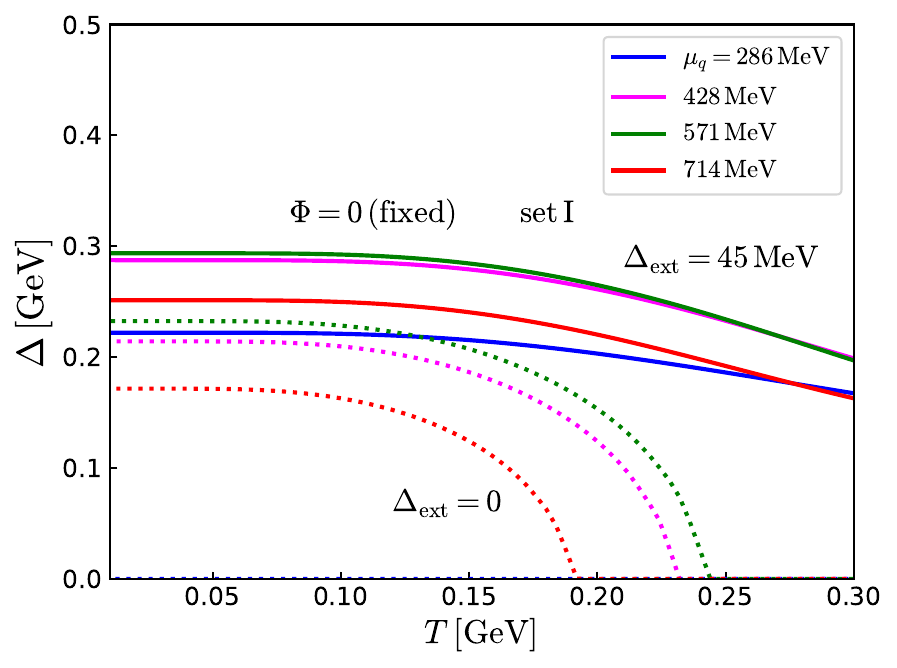}
\includegraphics[width=6.9 cm]{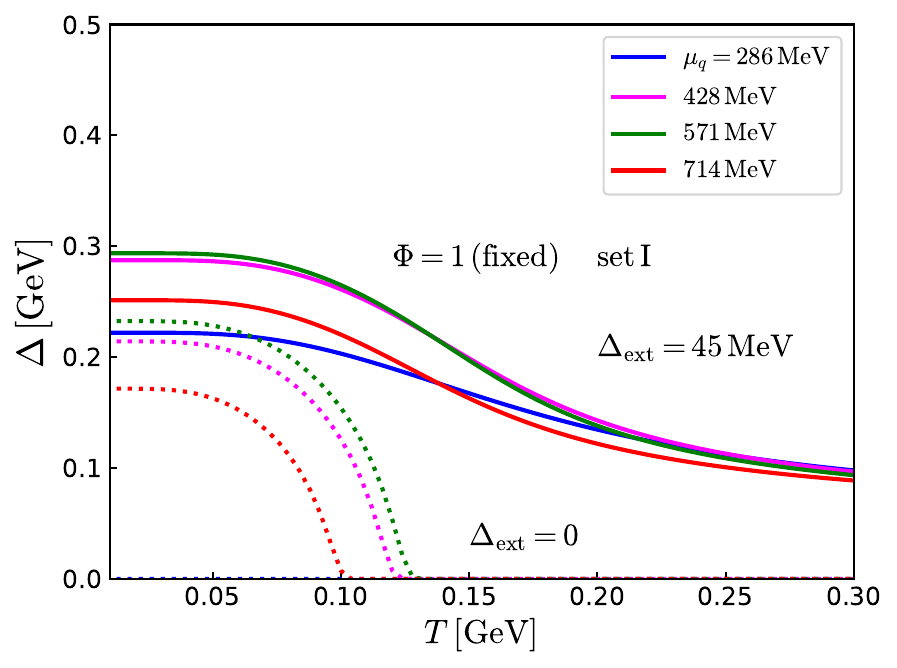}
\caption{ The same 
 plot as Figure~\ref{fig:Del_Tvary}, but $\Phi$ is fixed to 0 (\textbf{left panel}) or 1 (\textbf{right panel}).
}
\label{fig:Del_Tvary_Phi_fixed}
\end{figure}  

\begin{figure}[H]
\includegraphics[width=9 cm]{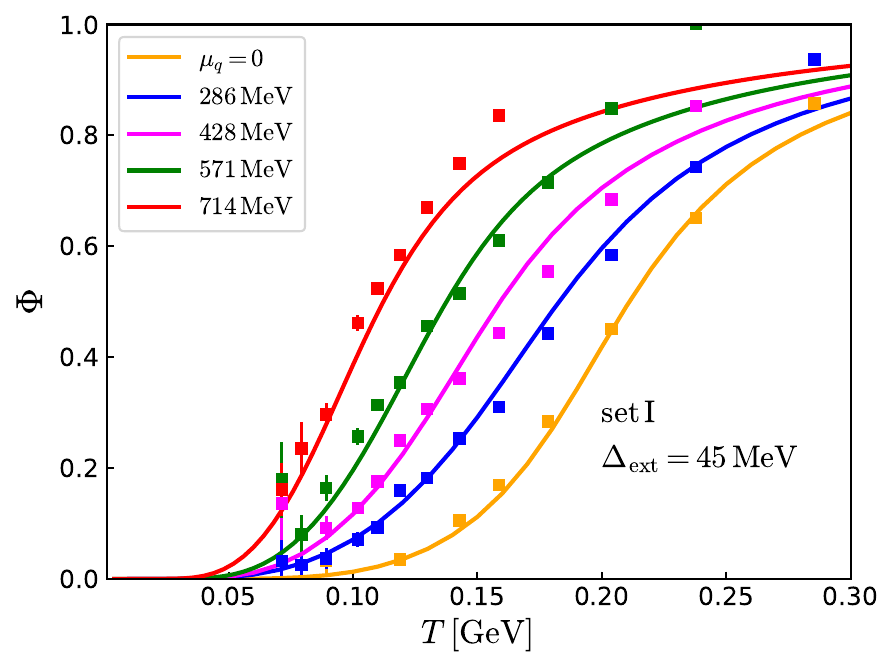}
\caption{ The temperature dependence of the Polyakov loop for set I at various $\mu_q$ with  \mbox{$\Delta_{\rm ext} = 45$~MeV.}
The lattice data is from the $ja=0.03$ in Ref.~\cite{Boz:2019enj}.
For the density interval from $\mu_q \simeq 300$ MeV to $\simeq 700$ MeV, 
the size of $\Delta$ is $\sim$250--300 MeV; see Figure~\ref{fig:gap_T0010}.
}
\label{fig:Phi_Tvary}
\end{figure}

Having established reasonable consistency in $T_s$ and $\Delta$ between model set I and the lattice results of Ref.~\cite{Boz:2019enj}, we now extend the comparison to finite $\mu_q$.
Figure~\ref{fig:Phi_Tvary} shows the Polyakov loop $\Phi$ for set I with $\Delta_{\rm ext}=45$ MeV, together with the lattice data at $ja =0.03$.
At low temperatures, corresponding to $\Phi \lesssim 0.8$, the~model reproduces the lattice results well.
At higher temperatures ($\Phi \gtrsim 0.8$), the~model $\Phi$ begins to saturate, while the lattice $\Phi$ continues to grow, as~the renormalization factor allows $\Phi$ to exceed unity.
This high-temperature deviation occurs for all $\mu_q$ studied, and~its origin remains unclear.
Nevertheless, our model is expected to be more reliable at low energies, and~the agreement in this regime is~noteworthy.
\vspace{-3pt}

\begin{figure}[H]
\includegraphics[width=6.9 cm]{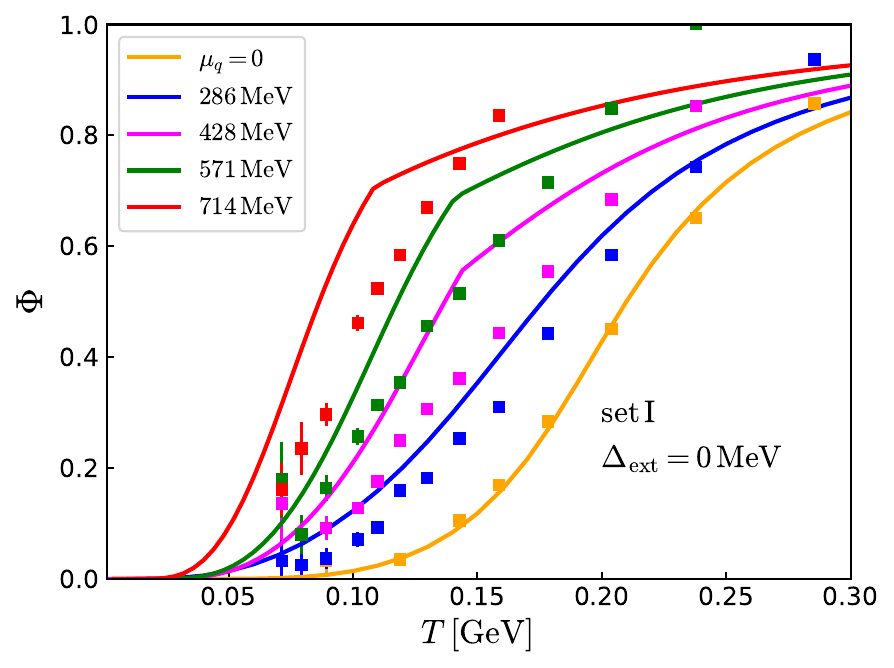}
\includegraphics[width=6.9 cm]{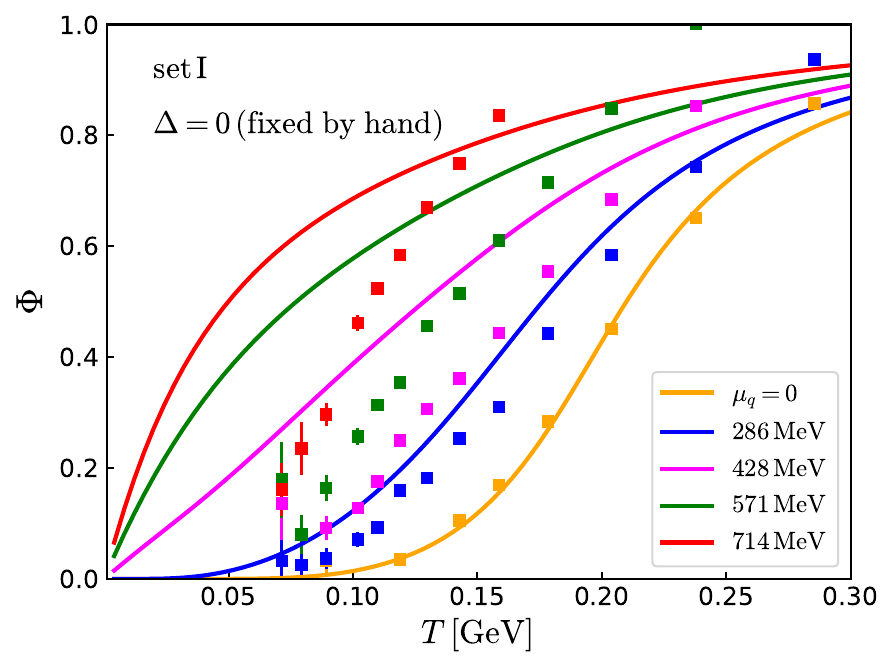}
\caption{ Same as Figure~\ref{fig:Phi_Tvary}, except: 
(\textbf{Left panel}) For the external diquark source, $\Delta_{\rm ex} = 0$ is chosen. 
Here, the phase transition from the diquark condensed phase to the normal phase is the second-order phase transition; 
(\textbf{Right panel}) $\Delta =0$ is fixed by hand for all $T$. The~gapless quarks trigger a rapid increase in the Polyakov loop already at temperatures less than $\sim$50 MeV.
}
\label{fig:Phi_Tvary_Del0}
\end{figure}

At low temperatures, it is crucial to maintain a sizable $\Delta$.
Figure~\ref{fig:Phi_Tvary_Del0} illustrates that if~$\Delta$ is artificially set to zero, quarks become gapless.
These gapless excitations induce a rapid rise of the Polyakov loop, resulting in substantial disagreement between our model and the lattice~data.

We emphasize that, in~all these analyses, the~parameters for the gluonic part of the Polyakov loop potential, $\Omega_{\rm pYM}$, are kept unchanged.
In principle, in-medium gluon screening could modify the gluon propagator and weaken confinement.
Nevertheless, our minimal setup—without major modifications to $\Omega_{\rm pYM}$—yields reasonable agreement with lattice results, 
suggesting that drastic changes in $\Omega_{\rm pYM}$ are disfavored for the density range considered here.
In the next section, we reinforce this conclusion by examining the effect of the diquark gap on the gluon~propagators.

\section{Gluon Propagators at One-Loop with Polyakov~Loops} \label{sec:gluons_one_loop_formula}

We focus on medium corrections to the static gluon propagators, $D(k)$, at~zero frequency ($k_0 = 0$).
Processes with $k_0 \ll \mu_q$ keep quarks near the Fermi surface and hence are most relevant for non-perturbative effects.
At the same time, the~momentum transfer $|\vk|$ can be large if quarks remain close to the Fermi surface before and after scattering.
Therefore, we concentrate on small $k_0$ while considering a wide range of $|\vk|$ to $\sim$$2\mu_q$.

To reproduce the lattice results of Ref.~\cite{Boz:2018crd}, it is crucial to suppress quark excitations in the infrared region.
Gapless quarks would otherwise over-screen the electric gluons~\cite{Contant:2019lwf}.
Including a diquark gap of $\Delta$$\sim$$200$ MeV reduces the electric screening, improving agreement with lattice data~\cite{Suenaga:2019jjv,Kojo:2021knn}.
In contrast, in~the magnetic sector, static screening remains absent for any value of $\Delta$, as~in the normal phase.
The magnetic screening is usually regarded as non-perturbative~effects.

At finite temperatures, thermally excited particles—either quarks or hadrons—respond to external momenta as gapless modes~\cite{Kojo:2021knn,Suenaga:2021bjz}.
If the thermal excitations are quark-dominated, they induce static electric screening as in the normal phase, while the static magnetic modes remain unscreened.
In contrast, if~the thermal excitations are confined, the~electric screening is expected to be weaker, since color-singlet objects of typical size $R_{\rm size}$ cannot respond effectively to external fields at wavelengths much longer than their size ($k \ll R_{\rm size}^{-1}$).

We account for the suppression of thermal quarks by including Polyakov loops.
The gluon self-energy is described by the polarization function
\beq
- \rmi \Pi_{\mu \nu}^{AB} (k_A) 
\equiv (\rmi g_s)^2 \int_{x,y} \rme^{-\rmi k(x-y) } 
	 \big\la \big( \bar{q} \gamma_\mu \frac{\, \lambda_A \,}{2} q \big) (x) \big(\bar{q} \gamma_\nu \frac{\, \lambda_B \,}{2} q \big)(y) \big\ra \,.
\label{eq:Pi_kA}
\eeq	 
\textls[-15]{where $k_A$ was defined in Equation~\eqref{eq:k_def}.
Explicit evaluation shows that $\Pi^{AB}$ depends on $k_A$ and is diagonal in color space, 
with nonzero components only for \mbox{$(A,B)=(3,3), (+,+), (-,-)$}.} All other combinations~vanish.

As usual, we decompose the gluon self-energy into electric and magnetic components.
In the presence of the background field, the~momenta entering the projection tensors are $k_A$,
which ensures the Ward--Takahashi identity, $k^\mu_A \Pi_{\mu \nu}^{AB} = 0$, is satisfied.
The decomposition then reads
\beq
\Pi^{AB}_{\mu \nu} (k_A) 
= \delta^{AB} P^E_{\mu \nu} (k_{A}) \Pi_E (k_{A}) + \delta^{AB}  P^M_{\mu \nu} (k_{A}) \Pi_M (k_{A}) 
\eeq
where $k_A$ was defined in Equation~\eqref{eq:k_def}.

\subsection{The Vacuum~Subtraction} \label{sec:vac_subtraction}

The polarization function is formally UV divergent.
We extract the UV divergent vacuum contribution and combine it with the counter terms for the renormalization,
\begin{align}
& \big[ \Pi_{\mu \nu}^{AB} (k_A) \big]^R
 \equiv \big[ \Pi_{\mu \nu}^{AB} (k_A) \big] + \delta Z_g k_A^2  P_{\mu \nu}^{k_A}
\notag \\
& = \delta^{AB} P_{\mu \nu}^{k_A} \big[ \Pi_{\rm vac} (k_A) +  \delta Z_g k_A^2 \big]
+ P^E_{\mu \nu} (k_{A}) \delta \Pi_E^{AB} (k_{A}) + P^M_{\mu \nu} (k_{A}) \delta \Pi_M^{AB} (k_{A})  \,,
\end{align}
where the renormalized vacuum part is ($K^2 \equiv - (k_A)^2$, $\alpha_s = g_s^2/4\pi$)
\vspace{6pt}
\beq
 \Pi_{\rm vac}^R (k_A) 
= \Pi_{\rm vac} (k_A) +  \delta Z_g k_A^2 
= - K^2 \frac{\, 2 \alpha_s^q \,}{\, \pi \,} 
\int_0^1 \rmd x \, x(1-x) \ln \frac{\, \big( M_q^{\rm vac} \big)^2 + x(1-x) K^2 \,}{\, \big (M_q^{\rm vac} \big)^2 + x(1-x) \mu_R^2 \,} \,,
\label{eq:vac_R}
\eeq
with $\mu_R$ being the renormalization scale.
The medium part is defined by subtracting the vacuum part,
\beq
\delta \Pi_{E,M} \equiv \Pi_{E,M} - \Pi_{\rm vac} \,,
\eeq
where the subscript $AB$ is omitted for the~moment.

This subtracted function is UV convergent, but~its finite part depends on regularization methods.
In particular, if~we integrate $q_0$ first to pick up poles and then integrate $\vq$,
there emerge UV finite terms which violate the conservation law $k_A^\mu \Pi_{\mu \nu}^{AB} (k_A) = 0$
when the quark propagators in $\Pi_{E,M}$ and $\Pi_{\rm vac}$ contain, e.g.,~different $M_q$ or $\Delta$ \cite{Kojo:2014vja}.

One way to restore the conservation law is to introduce counterterms that cancel regularization artifacts~\cite{Kojo:2014vja}.
Another approach is to split the vacuum subtraction into two steps~\cite{Kojo:2021knn}:
\beq
\delta \Pi_{E,M}
= \delta_{S} \Pi_{E,M} + \delta_{\mu_q} \Pi_{E,M}
= \big[ \Pi_{\rm vac}^{M_q\rightarrow \tilde{M}_q } - \Pi_{\rm vac} \big]_{\rm dim\,reg}
+ \big[ \Pi_{E,M} - \Pi_{\rm vac}^{M_q\rightarrow \tilde{M}_q } \big]_{\rm 3d\, reg}
 \,.
 \label{eq:dPi}
\eeq
Here,~$\delta_{\mu_q} \Pi_{E,M}$---which is the difference between $\Pi_{E,M}$ and $\Pi_{\rm vac}$ evaluated with the same gap parameters but different $\mu_q$---has no regularization artifacts;~hence, one can evaluate it using the standard residue method with a three-dimensional cutoff.
Meanwhile, each term in $\delta_{S} \Pi_{E,M}$ can be evaluated via dimensional regularization;~hence, $\delta_{S} \Pi_{E,M}$ can be made artifact-free.
The parameter $\tilde{M}_q (\mu_q, T) \equiv \sqrt{M_q^2 + \Delta^2}$, determined at a given $\mu_q$, 
is uniquely fixed to ensure that the conservation law is maintained.
Explicitly, $\delta_{S} \Pi_{E,M}$ reads as follows:
\beq
\delta_{S} \Pi_{E,M} 
= - K^2 \frac{\, 2 \alpha_s^q \,}{\, \pi \,} 
\int_0^1 \rmd x \, x(1-x) \ln \frac{\, \tilde{M}_q^2 + x(1-x) K^2 \,}{\, \big (M_q^{\rm vac} \big)^2 + x(1-x) K^2 \,} \,.
\eeq
If the gap parameters in vacuum and medium remain the same, i.e.,~$\tilde{M}_q = M_q^{\rm vac}$,
then $\delta_{S} \Pi_{E,M}$ vanishes by~construction.

\subsection{The Medium~Corrections} \label{sec:medium_correction}

Physical medium effects are largely reflected by $\delta_{\mu_q} \Pi_{E,M}$.
Decomposing the normal and anomalous quark propagators into the particle and antiparticle components, then calculating the residue integral for $q_0$, we find (before the vacuum subtraction)
\beq
- \rmi \Pi_{E, M}^{AB} (k_A) 
= \frac{\, g_s^2 \,}{2} \sum_{s,s'=p,a} \int_{\vq} 
K_{E,M}^{ss'} 
\big[ C_{E,M}^{ss'} \calG^{ss'}_{AB} 
+ \tilde{C}_{E,M}^{ss'} \tilde{\calG}^{ss'}_{AB} 
\big](\vq_+, \vq_-; k_0) 
 \,,
\eeq
where $\vq_\pm = \vq \pm \vk/2$.
We call $K_{E,M}^{ss'} $, $C_{E,M}^{ss'} $, and~$\calG^{ss'}_{AB}, \tilde{\calG}^{ss'}_{AB} $
the kinematic, coherence, and~propagator factors.
The propagator $\tilde{\calG}$ is non-vanishing only at nonzero~temperatures.

\subsubsection{Kinematic~Factors} \label{sec:kinematic}

The explicit forms of the kinematic factors are
\vspace{6pt}
\begin{eqnarray}
&& {\cal K}^{ \rmp \rmp}_{E} =  {\cal K}^{ \rma \rma}_{E}  = - \frac{k^2}{\vk^2} \bigg( 1 + \frac{\, \vec{q}^2 - \vk^2/4 + M_q^2 \,}{ E_{q_+}E_{q_-} } \bigg) \ , 
\nonumber \\
&& {\cal K}^{ \rmp \rma }_{E} =  {\cal K}^{ \rma \rmp}_{E}  =  - \frac{k^2}{\vk^2} \bigg(1 - \frac{\, \vec{q}^2 - \vk^2/4 + M_q^2 \,}{ E_{q_+}E_{q_-} } \bigg) \ , 
\nonumber \\
&& {\cal K}^{ \rmp \rmp}_{M} =  {\cal K}^{ \rma \rma}_{M}  = -1 + \frac{\, ( |\vec{q}|\cos\theta)^2 - \vk^2/4 + M_q^2 \,}{ E_{q_+}E_{q_-}}  \ ,
\nonumber \\
&& {\cal K}^{ \rmp \rma}_{M} =  {\cal K}^{ \rma \rmp}_{M}  = -1 - \frac{\, ( |\vec{q}|\cos\theta)^2 - \vk^2/4 + M_q^2 \,}{ E_{q_+}E_{q_-}}  \ ,
\end{eqnarray}
where $\theta$ in $\cos\theta$ is the angle between $\vq$ and $\vk$.
These kinematic factors depend on $M_q$ and the spatial momenta, and~they are common across different states of matter.
It is useful to note that, at~low momenta, the~electric components behave as
$ {\cal K}^{ \rmp \rmp}_{E} \rightarrow 2 + O(\vk^2)$ 
and ${\cal K}^{ \rmp \rma }_{E} \rightarrow O(\vk^2)$.
This indicates that in-medium corrections to the electric part are dominated by particle–hole~excitations.

\subsubsection{Coherence~Factors} \label{sec:kinematic}
The coherence factors, which characterize the diquark condensed phase, are
\begin{align}
{\cal C}^{ \rmp \rmp}_{E,M} &= \frac{1}{2} \left( 1 - \frac{\, ( E_{q_+} - \mu_q  )(  E_{q_-}  - \mu_q ) \pm |\Delta|^2 \,}{ \xi_\rmp^+ \xi_\rmp^-  } \right) \,, \nonumber\\
{\cal C}^{ \rma \rma}_{E,M} &= \frac{1}{2} \left( 1 - \frac{\, ( E_{q_+} +  \mu_q )( E_{q_-} + \mu_q ) \pm |\Delta|^2 \,}{ \xi_\rma^+ \xi_\rma^- } \right) \,, \nonumber\\
{\cal C}^{ \rmp \rma}_{E,M} &= \frac{1}{2} \left( 1 + \frac{\, ( E_{q_+} -  \mu_q )( E_{q_-} + \mu_q ) \mp |\Delta|^2 \,}{ \xi_\rmp^+ \xi_\rma^- } \right) \,, 
\end{align}
and the factor relevant only at finite temperature (since $\tilde{\calG}$ is nonzero only at finite temperature, see below),
\begin{align}
\tilde{\calC}^{ \rmp \rmp}_{E,M} &= \frac{1}{2} \left( 1 + \frac{\, ( E_{q_+} - \mu_q  )(  E_{q_-}  - \mu_q ) \pm |\Delta|^2 \,}{ \xi_\rmp^+ \xi_\rmp^-  } \right) \,, \nonumber\\
\tilde{\calC}^{ \rma \rma}_{E,M} &= \frac{1}{2} \left( 1 + \frac{\, ( E_{q_+} +  \mu_q )( E_{q_-} + \mu_q ) \pm |\Delta|^2 \,}{ \xi_\rma^+ \xi_\rma^- } \right) \,, \nonumber\\
\tilde{\calC}^{ \rmp \rma}_{E,M} &= \frac{1}{2} \left( 1 - \frac{\, ( E_{q_+} -  \mu_q )( E_{q_-} + \mu_q ) \mp |\Delta|^2 \,}{   \xi_\rmp^+ \xi_\rma^- } \right) \,, 
\end{align}
where ${\cal C}^{ \rmp \rma}_{E, M} (q_+, q_-) = {\cal C}^{ \rma \rmp}_{E,M} (q_-, q_+)$ and $\tilde{\calC}^{ \rmp \rma}_{E, M} (q_+, q_-) = \tilde{\calC}^{ \rma \rmp}_{E,M} (q_-, q_+)$.
It is important to note that, at~zero temperature, the~electric component is vanishing in the small momentum limit, ${\cal C}^{ \rmp \rmp}_{E}$$\sim$$O(\vq \cdot \vk)$.
Later, we see that the propagator factor $ \calG^{\rmp \rmp}_{E}$ remains finite for $\vk \rightarrow 0$ 
so that the product  $ {\cal C}^{ \rmp \rmp}_{E} \calG^{\rmp \rmp}_{E}$ vanishes, 
meaning that the electric screening is absent in the diquark condensed phase.
This situation differs from the normal phase
where the propagator exhibits a singularity $\vk$ as $\sim$$1/\vq \cdot \vk$ for small $|\vk|$ and hence makes the product 
$ {\cal C}^{ \rmp \rmp}_{E} \calG^{\rmp \rmp}_{E}$ finite. 
In this case, the Debye mass emerges, cutting off gluons in the IR.
In contrast, the~particle--hole contribution to the magnetic component is finite, ${\cal C}^{ \rmp \rmp}_{M}$$\sim$$\Delta^2/\xi_p^2$.

\subsubsection{Propagator~Factors} \label{sec:kinematic}

Finally, we examine the propagator factors where the effects of the Polyakov loop enter.
First, we generalize the Fermi--Dirac distribution for quarks as
\beq
n_F (x \pm \rmi Q/2)
=  \frac{\, 1 + \Phi \rme^{\beta x}  \,}{\, \rme^{2\beta x} + 2 \Phi \rme^{\beta x}  + 1 \,}
\pm \rmi \frac{\, \tilde{\Phi} \rme^{\beta x}  \,}{\, \rme^{2\beta x} + 2 \Phi \rme^{\beta x}  + 1 \,}
\equiv L_F^{\bf \Phi} (x) \pm \rmi \tilde{L}_F^{\bf \Phi} (x) \,,
\eeq
where ${\bf \Phi} = \Phi + \rmi \tilde{\Phi} = \cos (\beta \calR/2) + \rmi \sin (\beta \calR/2)$.

For deconfined quarks at $( \Phi, \tilde{\Phi}) =(1,0)$, 
this function is reduced to the usual Fermi--Dirac distribution, $L_F^{ {\bf \Phi}=1 }  (x) \rightarrow n_F (x)$ and $\tilde{L}^{ {\bf \Phi} = 1 }_F (x) \rightarrow 0$.
The low temperature behavior of the real part is $L_F^{ {\bf \Phi}=1 } (\xi_s)$$\sim$$\rme^{-\beta \xi_s}$.

For confined quarks at $( \Phi, \tilde{\Phi}) =(0,1)$, we find $L_F^{ {\bf \Phi} =\rmi }  (x) \rightarrow n_F (2x)$ and $\tilde{L}_F^{ {\bf \Phi} =\rmi }  (x) \rightarrow \rme^{\beta x} n_F (2x)$.
At low temperatures, their behaviors are $L_F^{ {\bf \Phi}=\rmi } (\xi_s)$$\sim$$\rme^{-2\beta \xi_s}$ and $\tilde{L}_F^{ {\bf \Phi} =\rmi } (\xi_s)$$\sim$$\rme^{-\beta \xi_s}$, respectively.
For low energy excitations, 
the real part exhibits a stronger suppression, $\sim$$\rme^{-2\beta \Delta}$, which can be interpreted as the Boltzmann factor for hadronic excitations.
Meanwhile, the~imaginary part is suppressed as $\sim$$e^{-\beta \Delta}$, reflecting quark excitations decaying into the colored~background.

Using these generalized distribution functions, the~$A=B=3$ components of the propagator factors are
\begingroup
\makeatletter\def\f@size{9}\check@mathfonts
\def\maketag@@@#1{\hbox{\m@th\normalsize\normalfont#1}}%
\begin{align}
\calG^{A=B=3}_{ss'} (q_+,q_- ; {\bf \Phi} ) 
&= 
 \big[ 1 - L_F^{{\bf \Phi}} (\xi_s^- ) - L_F^{{\bf \Phi}} (\xi_{s'}^+  ) \big]
\bigg(  \frac{\, 1 \,}{\,  k_0 + \xi_s^- + \xi_{s'}^+  \,} +  \frac{\, 1 \,}{\,  -k_0 + \xi_s^- + \xi_{s'}^+  \,} \bigg) \,,
 \notag \\
 \tilde{\calG}^{A=B=3}_{ss'} (q_+,q_-; {\bf \Phi}) 
&= \big[ L_F^{{\bf \Phi}} (\xi_s^- ) - L_F^{{\bf \Phi}} (\xi_{s'}^+  ) \big]
	\bigg(  \frac{\, 1 \,}{\, k_0 - \xi_s^- + \xi_{s'}^+  \,} + \frac{\, 1 \,}{\, - k_0 - \xi_s^- + \xi_{s'}^+  \,} \bigg) \,,
\end{align}
\endgroup
which contain only the real part of the thermal distribution functions.
On the other hand, the~$(A,B)=(\mp, \mp)$ components are
\begingroup
\makeatletter\def\f@size{9}\check@mathfonts
\def\maketag@@@#1{\hbox{\m@th\normalsize\normalfont#1}}%
\begin{align}
\calG^{\mp \mp}_{ss'} 
&= 
 \big[ 1 - L_F^{{\bf \Phi}} (\xi_s^- ) - L_F^{{\bf \Phi}} (\xi_{s'}^+  ) \big]
\bigg(  \frac{\, 1 \,}{\,  k_0 \mp \rmi \calR + \xi_s^- + \xi_{s'}^+  \,} +  \frac{\, 1 \,}{\,  - (k_0 \mp \rmi \calR) + \xi_s^- + \xi_{s'}^+  \,} \bigg) 
 \notag \\
&~~~~
 \pm \rmi \big[ \tilde{L}_F^{{\bf \Phi}} (\xi_s^- ) + \tilde{L}_F^{{\bf \Phi}} (\xi_{s'}^+  ) \big]
	\bigg(  \frac{\, 1 \,}{\, k_0 \mp \rmi \calR - \xi_s^- + \xi_{s'}^+  \,} - \frac{\, 1 \,}{\, - (k_0 \mp \rmi \calR) - \xi_s^- + \xi_{s'}^+  \,} \bigg) \,,
\notag \\
\tilde{\calG}^{\mp \mp}_{ss'} 
&= 
 \big[ L_F^{{\bf \Phi}} (\xi_s^- ) - L_F^{{\bf \Phi}} (\xi_{s'}^+  ) \big]
\bigg(  \frac{\, 1 \,}{\,  k_0 \mp \rmi \calR - \xi_s^- + \xi_{s'}^+  \,} +  \frac{\, 1 \,}{\,  - (k_0 \mp \rmi \calR) - \xi_s^- + \xi_{s'}^+  \,} \bigg) 
 \notag \\
&~~~~ 
\pm \rmi \big[ \tilde{L}_F^{{\bf \Phi}} (\xi_s^- ) + \tilde{L}_F^{{\bf \Phi}} (\xi_{s'}^+  ) \big]
	\bigg(  \frac{\, 1 \,}{\, k_0 \mp \rmi \calR - \xi_s^- + \xi_{s'}^+  \,} - \frac{\, 1 \,}{\, - (k_0 \mp \rmi \calR) - \xi_s^- + \xi_{s'}^+  \,} \bigg) \,,
\end{align}
\endgroup
which contain the real and imaginary parts of the thermal distribution functions.
It is important to note that $\calG^{AB}$ and $\tilde{\calG}^{AB}$ are functions of $k_A$;
this verifies Equation~\eqref{eq:Pi_kA}, where
we wrote $\Pi^{AB}$ as functions of $k_A$.
We also note that the imaginary parts of $\calG$ and $\tilde{\calG}$ are common,
${\rm Im} \calG^{\mp \mp}_{ss'} = {\rm Im} \tilde{\calG}^{\mp \mp}_{ss'} $.
These functions vanish at $(k_A)_0 = 0$, and~we will therefore not discuss them~further.

\subsubsection{The Full Expression After Adding the pYM Type~Self-Energy} \label{sec: pYM self}

Below, we examine the electric and magnetic propagators at $K^2=-k_A^2$.
It is known that the vacuum propagators in the pYM theories 
are well described by the massive gluon propagator, including the gluon one-loop polarization function,
$\Pi^G_{\rm vac} (k; m_g, \alpha_s^G, \mu_R)$, whose explicit expression is given by Equation~(17) in Ref.~\cite{Tissier:2011ey}.
Here, $\mu_R$ is the renormalization scale for which we take $\mu_R =1 $ GeV,
and $\alpha_s^G$ is the coupling in the gluon sector,
which generally differs from the quark--gluon coupling $\alpha_s^q$.
In the following, we begin with the vacuum propagators fitted to lattice data and 
then examine the impact of in-medium quark contributions.
The propagators are expressed as
\vspace{6pt}
\begingroup
\makeatletter\def\f@size{9}\check@mathfonts
\def\maketag@@@#1{\hbox{\m@th\normalsize\normalfont#1}}%
\beq
D_{\mu \nu}^{AB} (k_A) 
= D_E^{AB} P_{\mu \nu}^E +  D_M^{AB}  P_{\mu \nu}^M \,,
~~~~~~~~~~
D_{E,M}^{AB} (k_A) = \frac{\, \delta^{AB} \,}{\, K^2 + m_g^2 + \Pi_{\rm vac}^G + \Pi_{\rm vac}^R + \delta \Pi_{E,M} \,} \,,
\eeq
\endgroup
where $\Pi_{\rm vac}^R$ and $\delta \Pi_{E,M}$ are defined in Equations~\eqref{eq:vac_R} and \eqref{eq:dPi}.

\section{One Loop vs. Lattice Results: Screening~Masses} \label{sec:screening_masses}

From now on, we compare our computations with the lattice results of Ref.~\cite{Boz:2018crd}.
For concreteness, we focus on the data at $\beta = 2.1$, with~the inverse lattice spacing \mbox{$a^{-1} =1.41$~GeV,} the~pion mass $m_\pi \simeq 637$ MeV,
spatial size $L_s \simeq 2.21$ fm, and~temporal size $L_t$ corresponding to temperatures ranging from $T\simeq 44$ MeV to $353$ MeV.
This setup is very similar to the ``fine ensemble'' lattice data in Ref.~\cite{Boz:2019enj} for the Polyakov loop in the previous section, 
where $\beta \simeq 2.1$, $a^{-1}=1.43$ GeV, and~$m_\pi \simeq 637$ MeV.
In Ref.~\cite{Boz:2018crd},
the results at $T\simeq 44$ MeV are regarded as effectively zero temperature data.
Here, however, we explicitly treat them as $T = 44$ MeV results,
since analytic calculations at $T = 0$ and $T = 44$ MeV show considerable deviations at high~densities.

Unfortunately, the~lattice data at small momenta are sensitive to system size,
and the statistical errors are non-negligible.
Moreover, the~lattice propagators were not renormalized, so the overall normalization of the field amplitude is arbitrary.
To minimize both statistical and systematic uncertainties,
we normalize each propagator by its value at $K^2 = 0$ at a chosen reference point of $(\mu_q, T)$.

For the analytic results, we generally use $(\Phi, M_q, \Delta)$ obtained from our gap Equation~\eqref{eq:gap_eq} with $\Delta_{\rm ext} = 45$ MeV,
except for the $\mu_q = 0$ case, where the lattice computations set the external field to $\Delta_{\rm ext} = 0$.
Unless otherwise stated, we take $\alpha_s^G = 1.0$; 
variations in $\alpha_s^G$ can largely be compensated by adjusting $m_g$, for~which we test several~values.

\subsection{Density Evolution of the Screening Masses: Qualitative Guides from Fixed $\Phi$ and $\Delta$ }

To establish theoretical baselines, we first examine the screening masses with Polyakov loops set to $\Phi = 0$ (confined) 
and $\Phi = 1$ (deconfined) for various diquark gaps, $\Delta = 1, 100,$ and $200$ MeV. 
The quark mass $M_q$ is determined by solving the gap equations for each $(\Phi, \Delta)$ configuration. 
These results serve as guidelines for comparison with the lattice data.
%

We define the screening masses as follows:
\beq
m^2_{E/M} (\mu_q,T) \equiv 1/D_{E/M} (k=0;  \mu_q, T) \,.
\eeq
However, these expressions are not directly usable until a normalization is specified. 
To remove this arbitrariness and reduce statistical fluctuations in the lattice data, 
we normalize the screening masses by their values at zero density and "zero temperature" for each $\mu_q$:
%
%
\beq
\overline{m}_{E/M} (\mu_q, T)
\equiv  \frac{\, m_{E/M} (\mu_q,T) \,}{\, m_{E/M} (\mu_q^*, T^*) \,} \,,~~~~~~~~~ (\mu_q^*, T^* )=(0, 44)\,{\rm MeV} \,.
\eeq

\begin{figure}[H]
\includegraphics[width=7.2 cm]{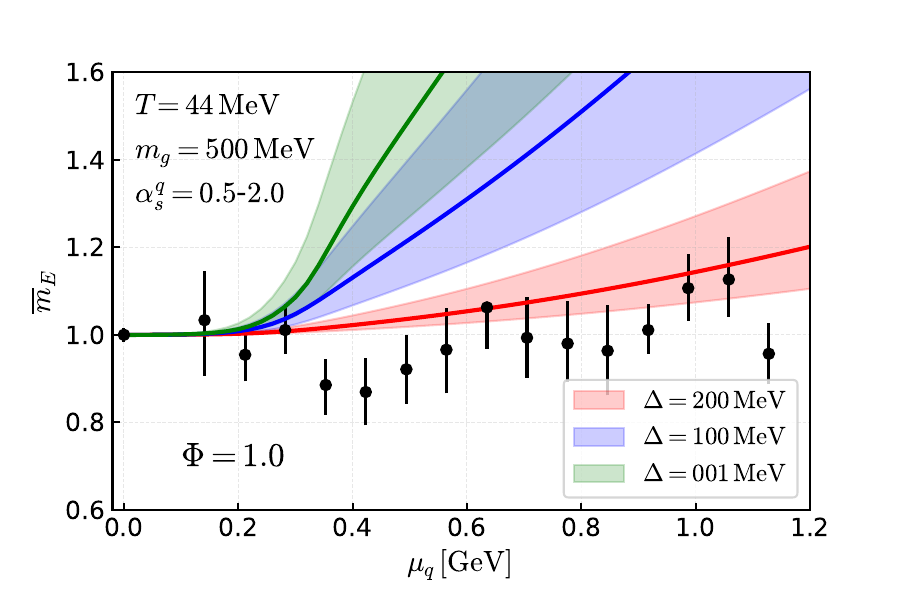}
\hspace{-0.7cm}
\includegraphics[width=7.2 cm]{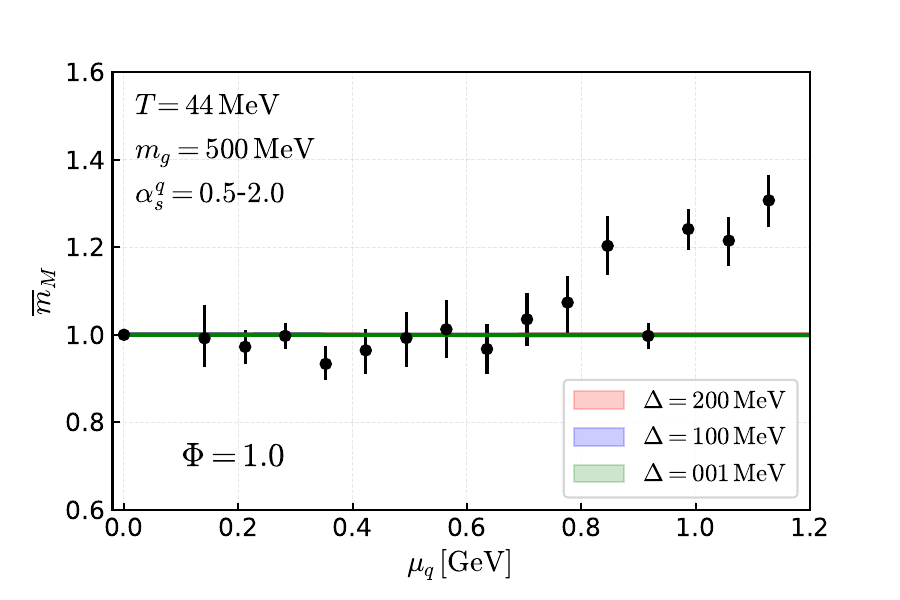}
\vspace{-0.55cm}
\\
\includegraphics[width=7.2 cm]{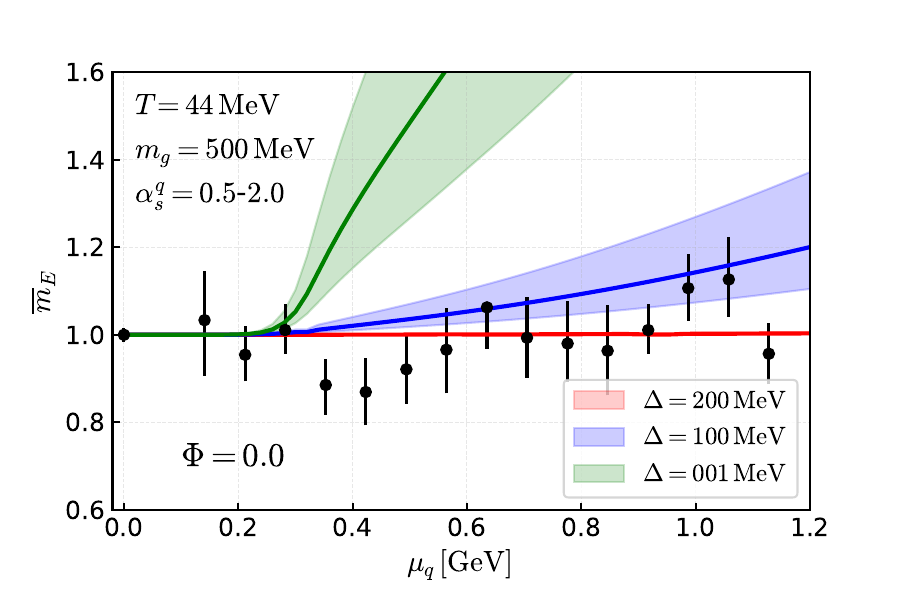}
\hspace{-0.7cm}
\includegraphics[width=7.2 cm]{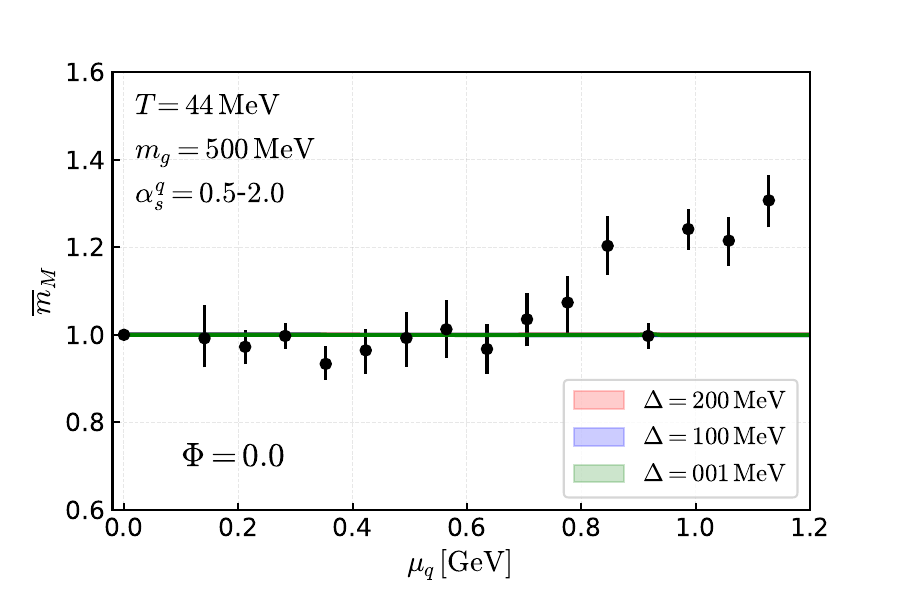}
\caption{Theoretical 
 guidelines vs lattice data for the normalized electric (\textbf{left panels}) and magnetic (\textbf{right panels}) screening masses, $\overline{m}_{E,M}$,
for the $\Phi=1$ (\textbf{upper panels}, deconfined) and $\Phi=0$ \mbox{(\textbf{lower panels}, confined)} cases.
The diquark gaps $\Delta=$ 1, 100, and~200 MeV and $m_g =500$ MeV are chosen for theoretical curves.
The error bands in theoretical curves originate from the variation in $\alpha_s^q$, ranging from 0.5 to 2,
with the central solid curve being $\alpha_s^q=1$. 
To fit the lattice data in the electric sector, it is essential to quench thermal quarks by the diquark gap and/or Polyakov loop effects;
even at temperatures as low as $T\simeq 44$ MeV, the~abundance of thermal quarks is not negligible at large densities.
In contrast, the~theory predicts that the magnetic sector is free from the quark \mbox{screening effects.}
}
\label{fig:screening_mass_T044}
\end{figure}

Shown in Figure~\ref{fig:screening_mass_T044} are the normalized electric (left panels) and magnetic (right panels) screening masses at low temperature, $T\simeq 44$ MeV.
For the analytic results, we display two cases: $\Phi = 1$ (upper panels, deconfined) and $\Phi = 0$ (lower panels, confined).
The error bands on the theoretical curves correspond to variations of $\alpha_s^q$ from 0.5 to 2, with~the central solid curve representing $\alpha_s^q = 1$.

For the electric sector, it is essential to consider a relatively large gap, $\Delta$$\sim$$200$ MeV, and/or to include the Polyakov loop suppression of thermal quarks.
The theoretical curves for the electric screening mass suggest that, for~$\Delta \lesssim 100$ MeV, the~screening masses increase too rapidly once the BCS regime sets in.
This estimate of $\Delta$$\sim$$200$ MeV is consistent with the lattice results for the superfluid critical temperature, $T_s \simeq 100$ MeV, in~the BCS region;
applying the BCS estimate, $\Delta^{\rm BCS} \simeq T_s / 0.57$$\sim$$180$ MeV~\cite{Pisarski:1999bf}, supports this~observation.

Including the Polyakov loop suppresses thermal quark contributions, tempering the evolution of the electric screening toward high densities.
Near the Fermi surface, the~abundance of thermal quarks and baryons (diquarks) scales as 
$\sim$$\Phi e^{-\beta \xi_p}$$\sim$$\Phi e^{-\beta \Delta}$ and $\sim$$e^{-2 \beta \Delta}$, respectively, controlling the strength of quark screening effects.
The impact of the Polyakov loop is small for $\Delta \simeq 0$ as $\rme^{-\beta \Delta} \simeq \rme^{-2\beta \Delta} \simeq 1$, while for a substantial $\Delta$, 
roughly twice the temperature is required for the confined case ($\Phi=0$) to develop a screening mass as large as in the deconfined case ($\Phi=1$).
Equivalently, for~a fixed $T$, the~results for $\Delta$ with $\Phi=0$ resemble those for $2\Delta$ with $\Phi=1$.
This correspondence can be verified by comparing the curve for $\Delta = 100$ MeV with $\Phi=0$ to that for $\Delta = 200$ MeV with $\Phi=1$,
i.e., $\rme^{- 2\beta \times 100\, {\rm MeV} } = \rme^{- \beta \times 200\, {\rm MeV} }$,
as shown in Figure~\ref{fig:screening_mass_T044}.
\vspace{-10pt}

\begin{figure}[H]
\includegraphics[width=7 cm]{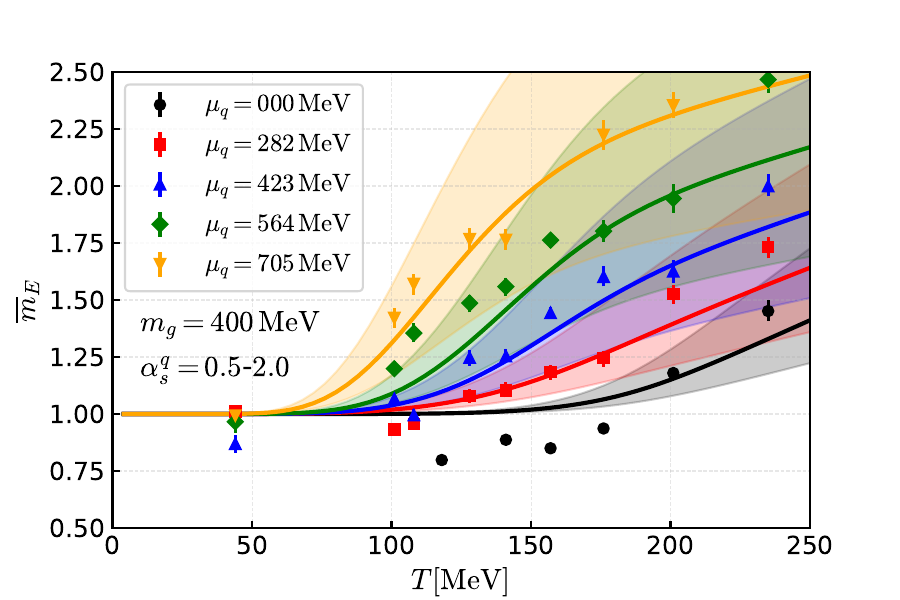}
\hspace{-0.7cm}
\includegraphics[width=7 cm]{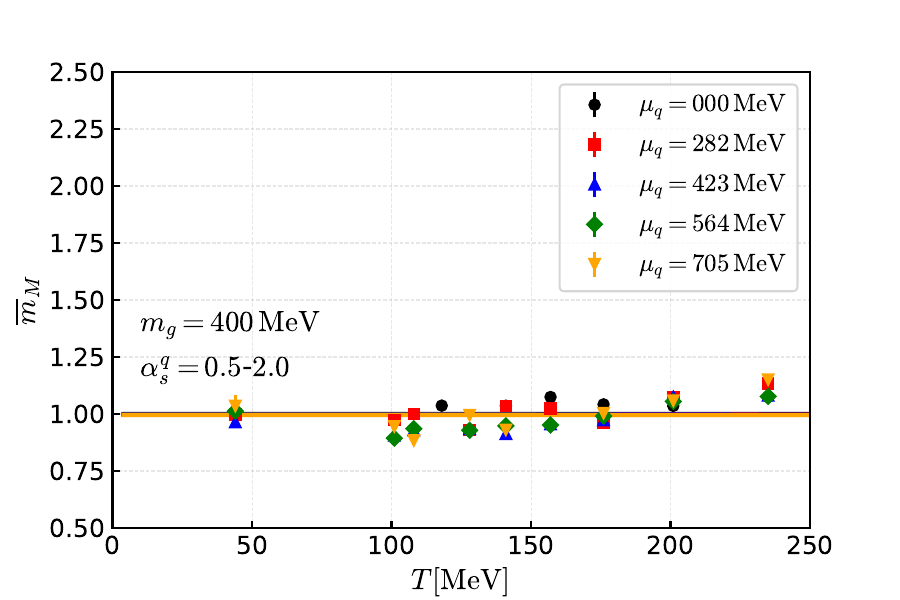}
\\
\includegraphics[width=7 cm]{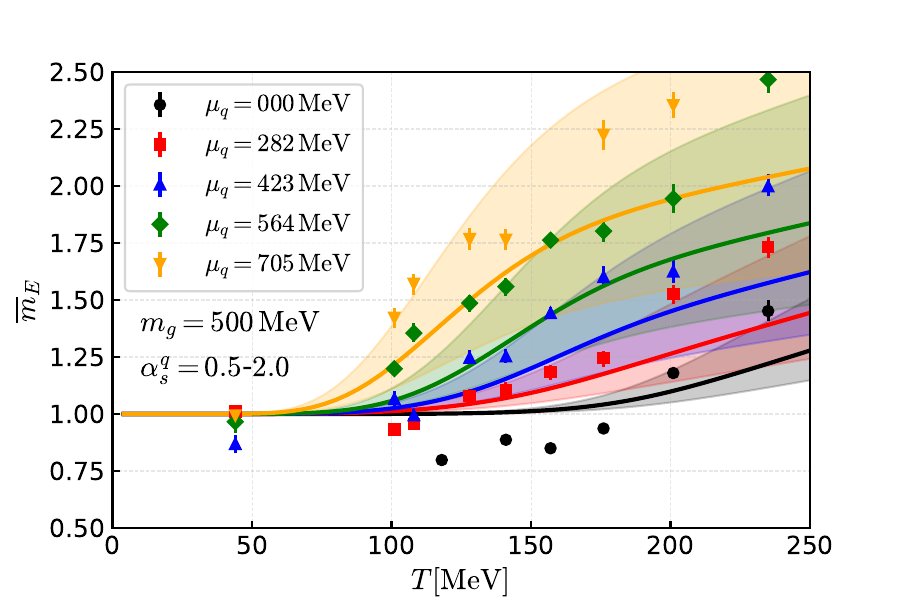}
\hspace{-0.7cm}
\includegraphics[width=7 cm]{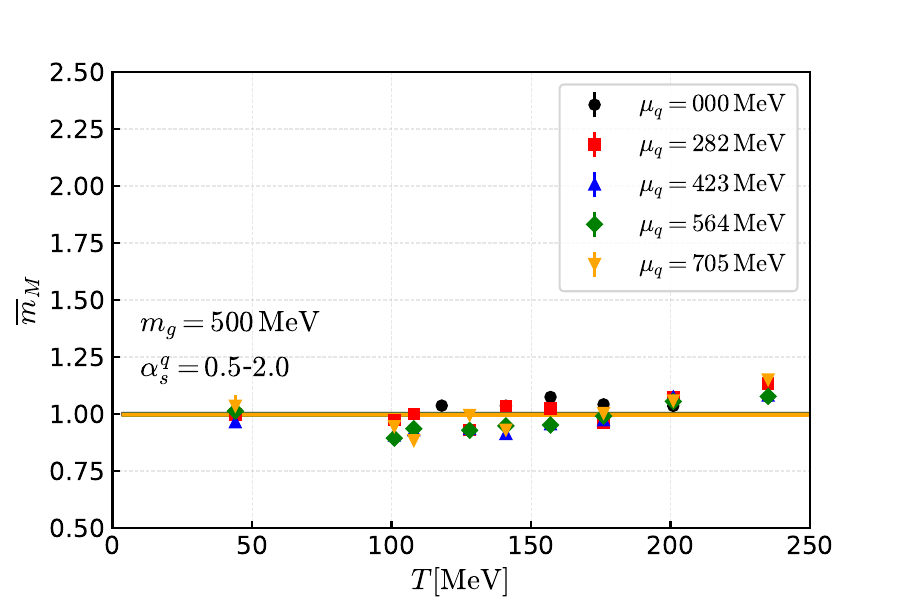}
\caption{The thermal evolution of the normalized electric (\textbf{left panels}) and magnetic (\textbf{right panels}) screening masses $\overline{m}_{E,M}$ for several $\mu_q$.
The cases for $m_g =$ 400 MeV (\textbf{upper panels}) and 500 MeV (\textbf{lower panels}) are displayed.
The error bands are set in the same way as Figure~\ref{fig:screening_mass_T044}.
}
\label{fig:screening_mass_global}
\end{figure}

The situation is very different in the magnetic sector.
The theoretical curves show essentially no in-medium modifications of the magnetic screening mass with increasing $\mu_q$ (and also with increasing $T$, see Figure~\ref{fig:screening_mass_global}).
The lattice results, however, suggest that the magnetic mass starts to increase from $\mu_q$$\sim$$700$ MeV,
a trend that cannot be captured by the present one-loop~calculations.

The values of $(\Phi, \Delta)$ dynamically determined from the gap equations remain within the range $\Phi$$\sim$$0$ and $\Delta$$\sim$$200$–300 MeV.
Consequently, the~screening masses closely follow the curves for $\Delta = 200$ MeV and $\Phi = 0$,
where both the electric and magnetic masses remain nearly constant over the range of $\mu_q$ explored in this work.
For this reason, we do not separately display the results of dynamically determined gaps~here.



\subsection{Thermal Evolution of the Screening~Masses}

Next, we examine the temperature evolution
of the static screening masses for $\mu_q =$ 0, 282, 423, 564, and~705 MeV.
Shown in Figure~\ref{fig:screening_mass_global}
are the normalized electric and magnetic screening masses.
For the analytic results, 
we present two cases with $m_g =$ 400 MeV (upper panels) and 500 MeV (lower panels).
The error bands for the theoretical curves reflect the variation in $\alpha_s^q$ from 0.5 to 2,
with the central solid curve corresponding to $\alpha_s^q = 1$.

For the electric sector, the~screening mass increases with $T$, as~expected from the conventional picture of the thermal Debye mass.
The electric screening is primarily governed by the gap $\Delta$; in the absence of $\Delta$, the~electric mass of order $\sim$$g_s \mu_q$ already emerges at $T=0$.
The Polyakov loop suppresses thermal quark fluctuations and thereby moderates the growth of thermal contributions to the screening.
The results with $m_g = 400$ MeV appear to reproduce the lattice data well, whereas those with $m_g = 500$ MeV tend to slightly underestimate them.
At present, however, it is difficult to pin down the optimal value of $m_g$ from these computations, since its effect can be compensated by the behavior of $\Delta$, which is not yet conclusively determined.
In fact, our determination of $\Delta$ is not entirely free from cutoff artifacts, as~already mentioned.
We suspect that the value of $\Delta$ in our PNJL model—estimated as $\simeq$$220$–$300$ MeV 
for $\Delta_{\rm ex} = 45$ MeV—and the resulting superfluid critical temperature $T_c$$\sim$$150$ MeV may be somewhat too large.
If this expectation is correct, the~quark contributions to $m_g$ should be more pronounced than in the present computations, 
in which case a larger $m_g$ or $\alpha_s^q$ value may be~acceptable.

For the magnetic sector, the~quark contributions to the screening mass vanish for all $\mu_q$ and $T$.
Consequently, the~magnetic screening mass remains insensitive to variations in $\mu_q$ and $T$.
This trend is consistent with the lattice data up to $T \simeq 250$ MeV.
At zero momentum ($|\vk|=0$), we find no in-medium modifications for the analytically computed magnetic propagators,
whereas at finite spatial momenta such modifications do appear, as~will be discussed in the next~subsection.

\section{One Loop vs. Lattice Results: Gluon Propagators at Finite~Momenta} \label{sec:gluons_at_finite_momenta}

Next, we examine the momentum dependence of the gluon propagators.
We focus on the static case with $k_0=0$ but cover a wide range of spatial momenta,
since such modes play the dominant role in low-energy processes.
We select specific $(\mu_q, T)$ points where lattice data are available~\cite{Boz:2018crd}.
Our analysis begins with the thermal evolution at $\mu_q=0$ and $\mu_q=705$ MeV,
followed by the density evolution at $T=44$ MeV and $T=118$ MeV.
The results at $T=118$ MeV, however, turn out to be very sensitive to the location of the diquark critical temperature
and therefore require some additional~analysis.

\subsection{Thermal Evolution at $\mu_q=0$ and $\mu_q=705$ MeV}

\begin{figure}[H]
\includegraphics[width=7. cm]{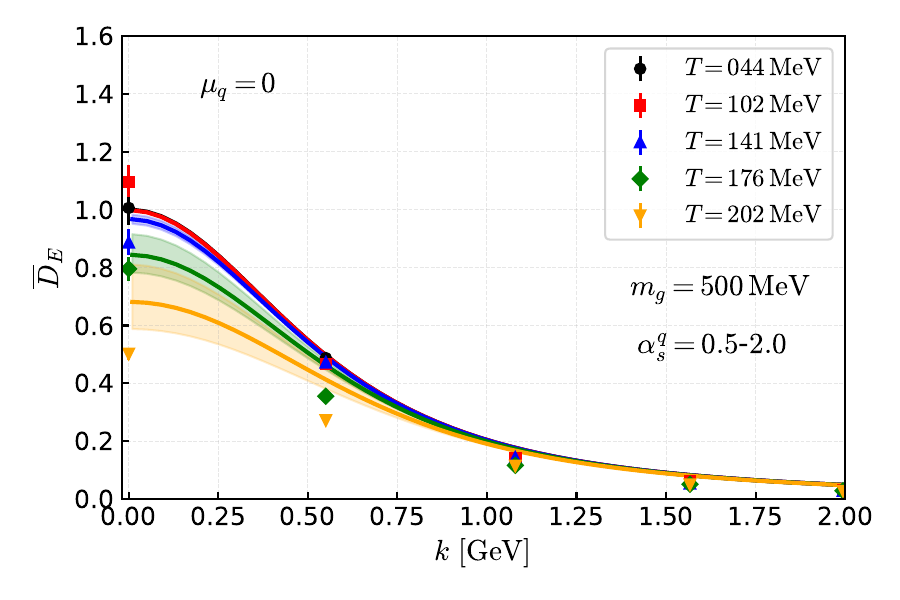}
\hspace{-0.5cm}
\includegraphics[width=7. cm]{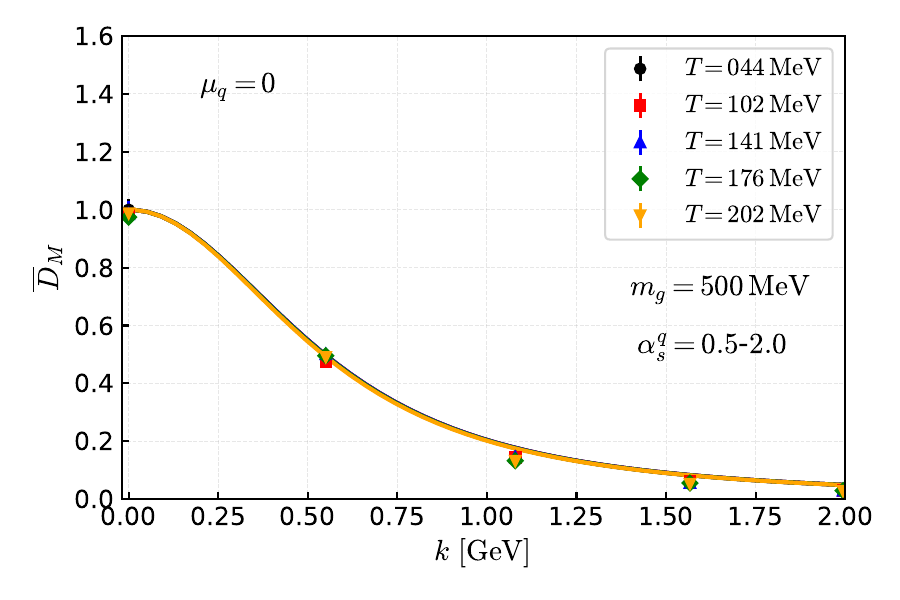}
\vspace{-.1cm}
\\
\includegraphics[width=7. cm]{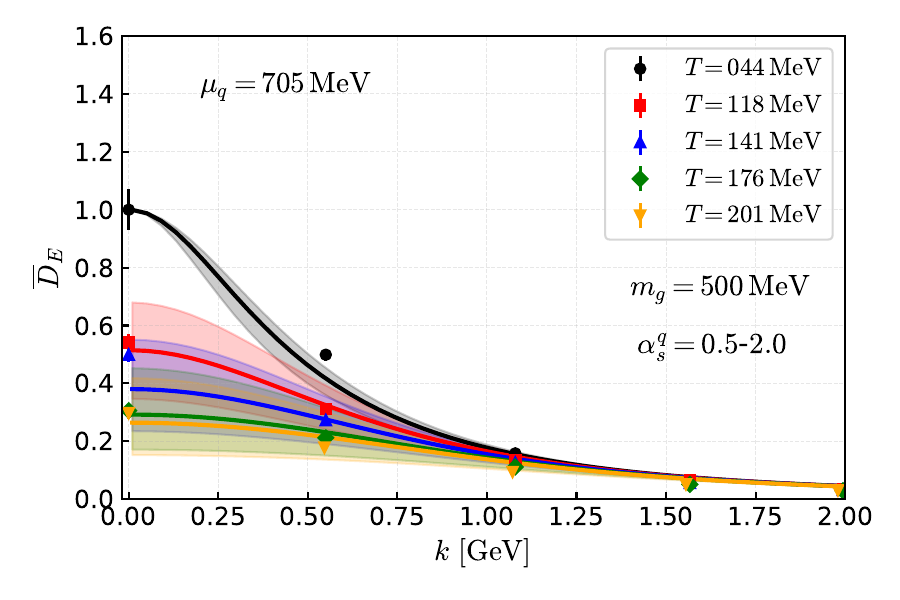}
\hspace{-0.5cm}
\includegraphics[width=7. cm]{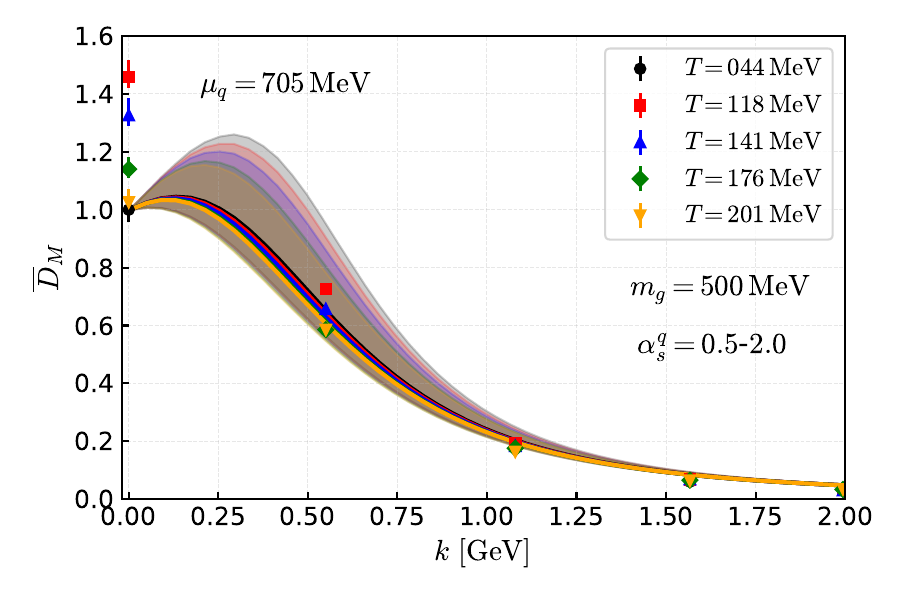}
\caption{The thermal evolution of the normalized electric and magnetic propagators $\overline{D}_{E,M} (k_0=0, \vk)$ for $\mu_q=0$ (\textbf{upper panels}) and $\mu_q=705$ MeV (\textbf{lower panels}).
The value of $m_g$ is fixed to 500 MeV, and the error bands are set in the same way as Figure~\ref{fig:screening_mass_T044}.
The magnetic propagators show the paramagnetic enhancement at finite momenta.
}
\label{fig:gluons_mu000_mu705}
\end{figure}

First, we examine the thermal evolution of the propagators at $\mu_q = 0$ and\mbox{ $\mu_q = 705$~MeV.}
Figure~\ref{fig:gluons_mu000_mu705} compares our analytic results with the lattice data.
The propagators are normalized as follows:
\beq
\overline{D}_{E/M} (\mu_q, T)
\equiv  \frac{\, D_{E/M} (\mu_q,T) \,}{\, D_{E/M} (\mu_q, T^*) \,} \,,~~~~~~~~~ T^* = 44 \,{\rm MeV} \,.
\eeq
Here, we show the results for $m_g = 500$ MeV,
with the range of $\alpha_s^q$ varied between 0.5 and 2.0, as~in the previous~analyses.
\vspace{-6pt}

For $\mu_q = 0$, the~electric propagators in the lattice data show a slightly stronger suppression than in our analytic results.
This discrepancy becomes less pronounced if we reduce the value of $m_g$, so we consider it to lie within the range of theoretical uncertainties.
The magnetic propagators, on~the other hand, are well reproduced across the entire temperature range~studied.

Next, we turn to the $\mu_q = 705$ MeV case.  
The electric propagators show good consistency between the analytic and lattice results.
In contrast, the~magnetic propagators exhibit a sizable discrepancy at $|\vk| = 0$, for~which we currently have no clear explanation.
This suggests that computations beyond one-loop order or genuinely non-perturbative effects may be required.
Meanwhile, the~analytic and lattice results are in good agreement for $|\vk| \gtrsim 500$ MeV.

\subsection{Density Evolution at $T=44$ MeV and $T=118$ MeV}

\begin{figure}[H]
\includegraphics[width=7. cm]{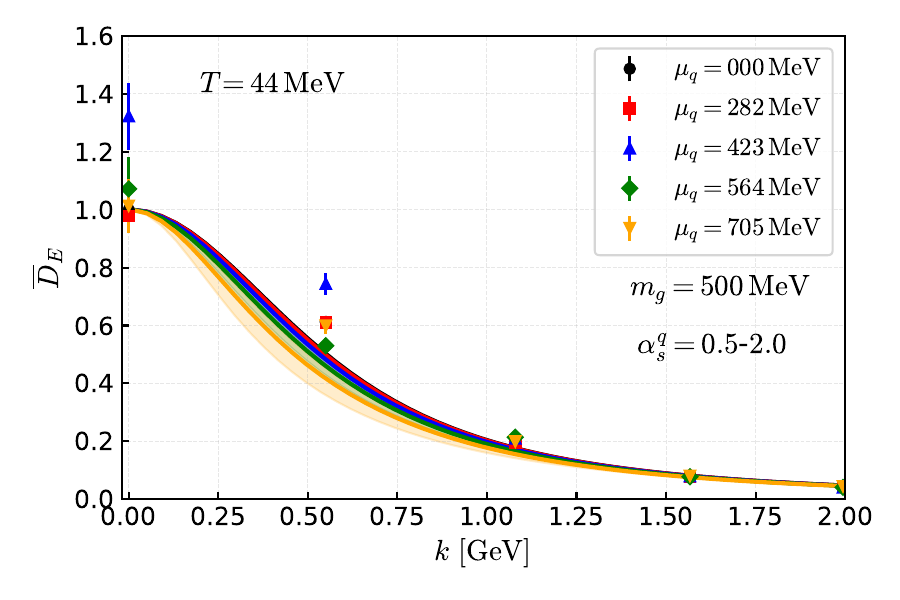}
\hspace{-0.5cm}
\includegraphics[width=7. cm]{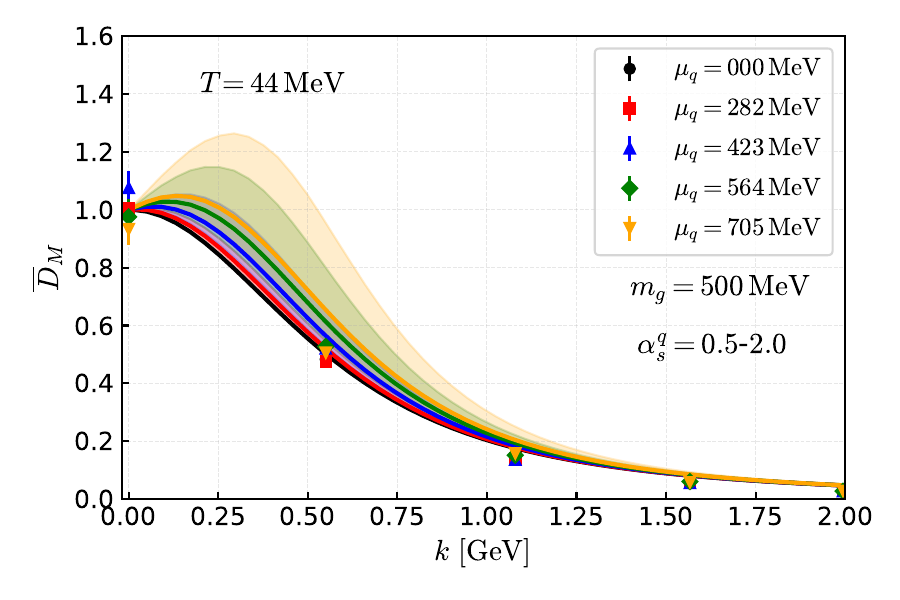}
\vspace{-.1cm}
\\
\includegraphics[width=7. cm]{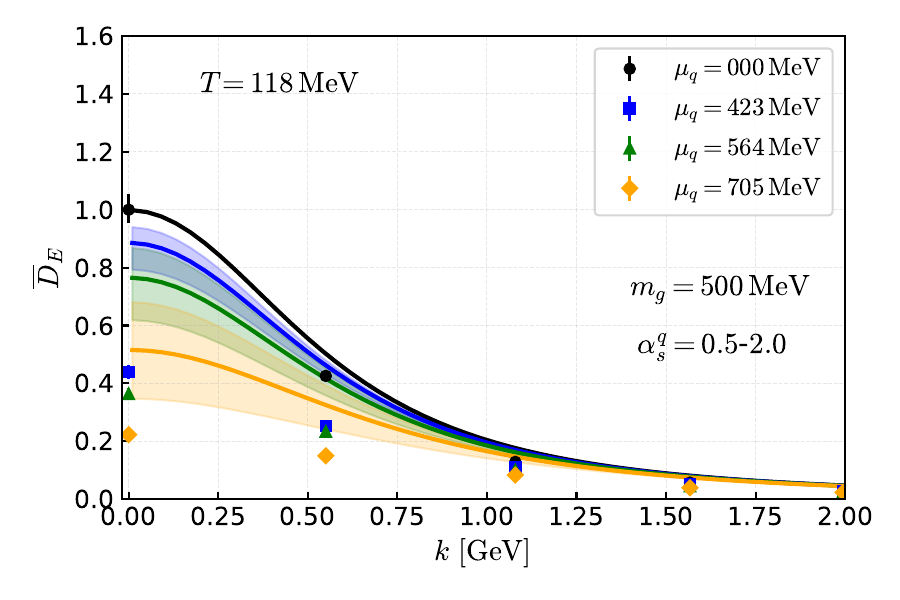}
\hspace{-0.5cm}
\includegraphics[width=7. cm]{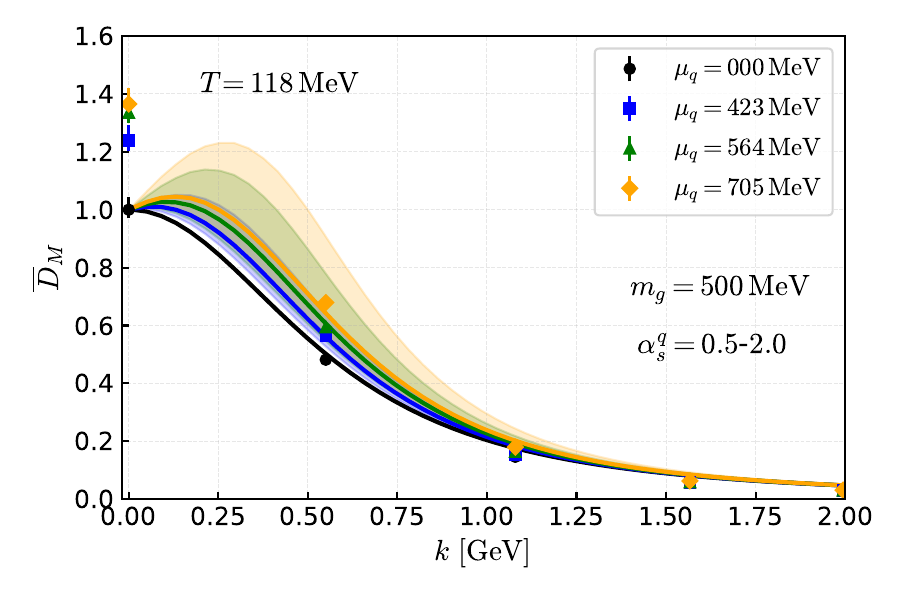}
\caption{The density evolution of the normalized electric and magnetic propagators \mbox{$\overline{D}_{E,M} ((k_0=0, \vk)$} for $T=44$ MeV \textbf{(upper panels)} and $T=118$ MeV (\textbf{lower panels}).
The value of $m_g$ is fixed to 500 MeV, and the error bands are set in the same way as Figure~\ref{fig:screening_mass_T044}.
The magnetic propagators show the paramagnetic enhancement at finite momenta.
At $T=118$ MeV, the $T$ dependence of the electric propagators  is considerably  weaker than the lattice data,
likely because $\Delta_{T=118\,{\rm MeV}}$ \mbox{is overestimated.}
}
\label{fig:gluons_T044_and_T118}
\end{figure}

Here, we examine the density evolution of the propagators at fixed temperatures, $T=44$ MeV and $T=118$ MeV.
Shown in Figure~\ref{fig:gluons_T044_and_T118} are comparisons between the analytic results and lattice data.
Here, the propagators are normalized as follows:
\beq
\overline{D}_{E/M} (\mu_q, T)
\equiv  \frac{\, D_{E/M} (\mu_q,T) \,}{\, D_{E/M} (\mu_q^*, T) \,} \,,~~~~~~~~~ \mu_q^* = 0 \,.
\eeq

First, we consider the $T=44$ MeV results.
In the lattice data, the~electric propagators at $\mu_q = 423$ and $564$ MeV show an enhancement at low momenta,
but we suspect that this feature is simply due to statistical fluctuations (see also Figure~\ref{fig:screening_mass_T044}).
For $\mu_q = 282$ and $705$ MeV, the~analytic results are consistent with the lattice data at $|\vk|$$\sim$$0$
but remain slightly smaller than the lattice values for $|\vk| \gtrsim 500$ MeV.
This tension can be alleviated by increasing $m_g$.
Taking these observations together, we conclude that the agreement between the analytic and lattice results at $T=44$ MeV is~acceptable.
\vspace{-3pt}

As for the magnetic propagators at finite momenta, we observe a significant quark contribution through paramagnetic enhancement, which grows with increasing $\mu_q$.
It is important to note that choosing too large a value of $\alpha_s^q$ leads to a serious disagreement between the analytic and lattice results.
Furthermore, extrapolating the one-loop results with fixed $\alpha_s^q$ to very high density eventually causes chromo-magnetic instability~\cite{Huang:2004bg,Huang:2004am}, which has been extensively discussed in the context of gapless color \mbox{superconductivity~\cite{Shovkovy:2003uu,Huang:2003xd}}.
A possible way to avoid such an instability is to invoke a “gluon saturation” mechanism, 
as in the Gribov–Zwanziger model~\cite{Gribov:1977wm,Vandersickel:2012tz}, 
which would drive $m_g$ to increase with $\mu_q$ before the paramagnetic enhancement causes the instability.
This conceptual issue, however, lies beyond the scope of the present~paper.

\begin{figure}[H]
\includegraphics[width=7. cm]{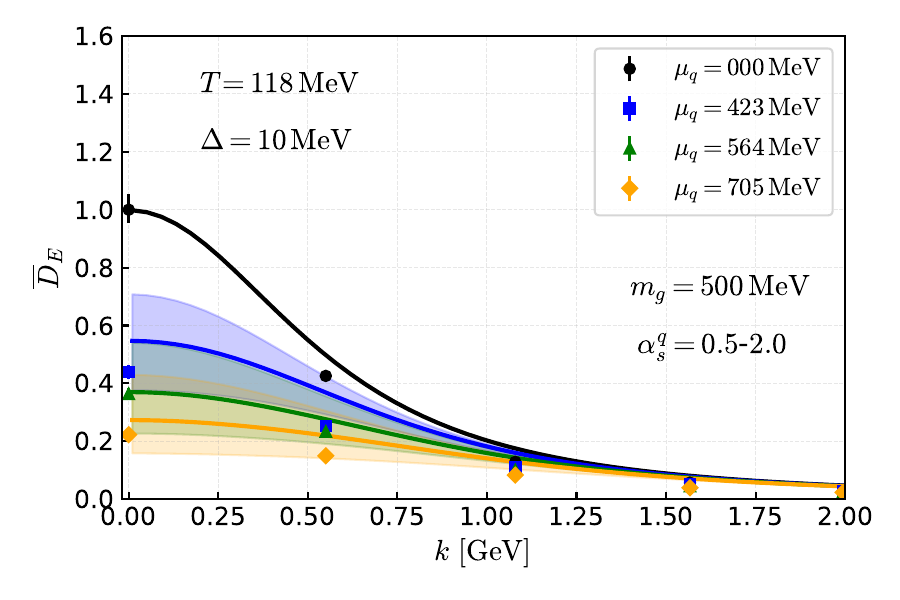}
\hspace{-0.5cm}
\includegraphics[width=7. cm]{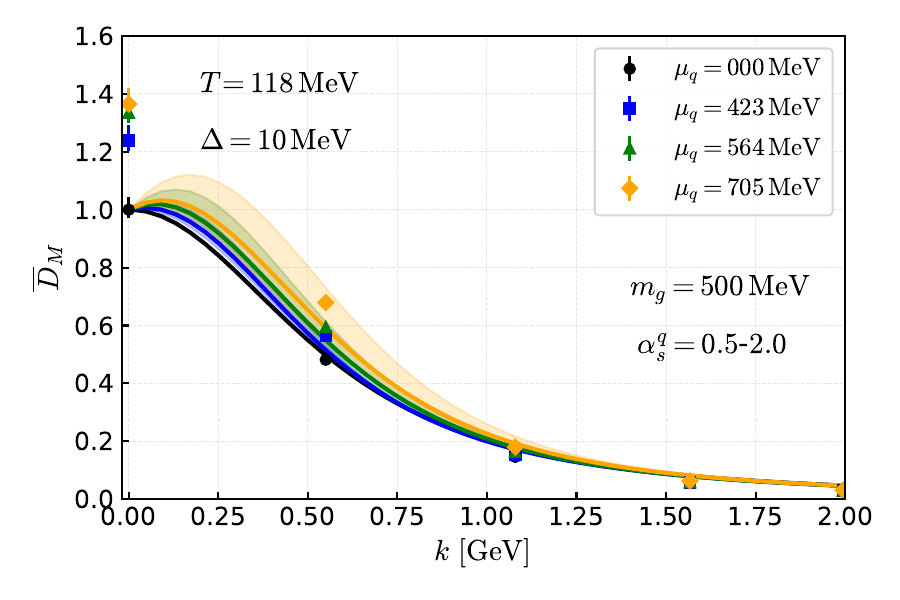}
\\
\includegraphics[width=7. cm]{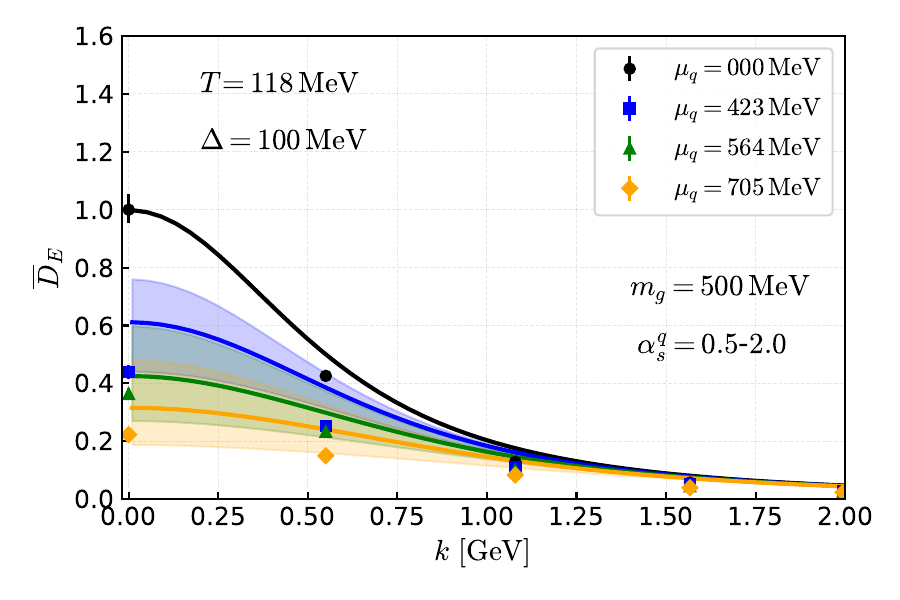}
\hspace{-0.5cm}
\includegraphics[width=7. cm]{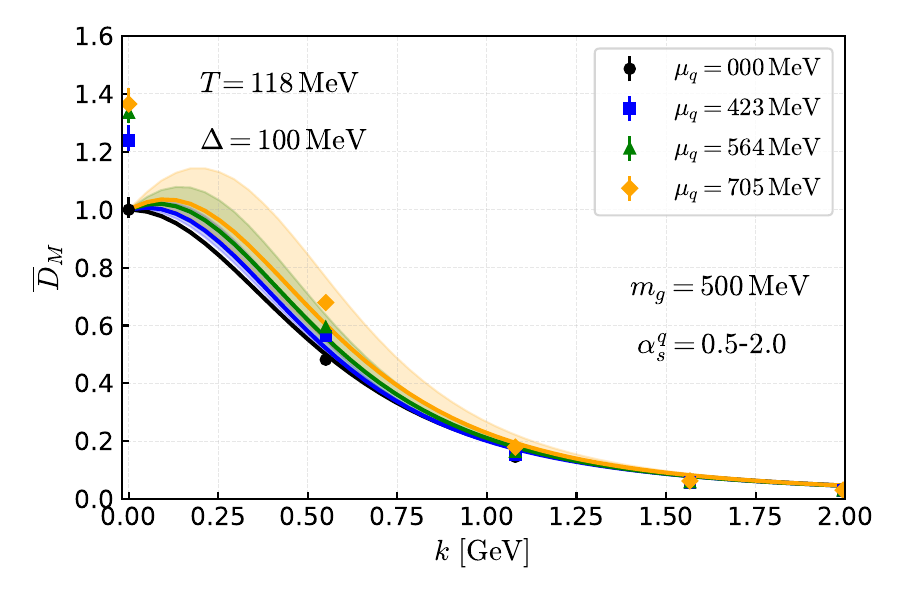}
\caption{The normalized electric and magnetic propagators $\overline{D}_{E,M} ((k_0=0, \vk)$ at $T=118$ MeV.
The diquark gaps are held fixed to $\Delta_{T=118} = 10$ MeV (\textbf{upper panel}) and $100$ MeV (\textbf{lower panel}).
}
\label{fig:muvary_T118_fixed_T_fixed_Del}
\end{figure}

Now, we turn to the case of $T=118$ MeV.
Our analytic results show substantially weaker electric screening compared to the lattice results.
This discrepancy may be understood as follows.
As mentioned earlier, the~$\Delta$ and the superfluid critical temperature $T_s$ in our PNJL model are about 1.2–1.5 times larger than those inferred from the lattice.
Consequently, while $T=118$ MeV lies slightly above the lattice $T_s$, it is still smaller than our model’s $T_s$ by about 20–50 MeV.
To examine how this discrepancy affects the results, we compute the analytic propagators with fixed values of $\Delta$ of $\Delta=10$ and $100$ MeV 
while determining $(\Phi, M_q)$ self-consistently from the gap equation.
The corresponding results are shown in Figure~\ref{fig:muvary_T118_fixed_T_fixed_Del}.
Using smaller $\Delta$ than predicted in our PNJL model enhances electric screening and hence suppresses the propagators, 
thereby improving agreement with the lattice~data.

For the magnetic propagators at $T=118$ MeV,
the agreement with the lattice results is generally good,
as shown in Figures~\ref{fig:gluons_T044_and_T118} and \ref{fig:muvary_T118_fixed_T_fixed_Del}.
We also notice that reducing $\Delta$ slightly weakens the paramagnetic enhancement at finite $|\vk|$.
\vspace{-3pt}

\section{Discussion and~Summary} \label{sec:summary}

In this paper, we delineate the role of thermal quarks in dense QC$_2$D and their impact on gluon dynamics, 
focusing specifically on the Polyakov loop and gluon~propagators.

Our initial approach was to start from the massive gluon propagators constrained by lattice data and then use them to compute the Polyakov loop potential.
However, the~resulting potential for the pYM turns out to be too flat; 
the inclusion of quark contributions drastically deforms the structure of the Polyakov loop potential, 
leading to an overly rapid deconfinement that disagrees with lattice QCD results.
We therefore speculate that the massive gluons appearing in the propagators and those responsible for confinement (the Polyakov loop potential) are somewhat distinct: 
the former corresponds to excitations above the confining vacuum, 
while the gluons forming the Polyakov loop potential are not like quasi-particles and appear to be more robust against quark-induced~perturbations.

To investigate the nature of thermal excitations in dense matter, we employ phenomenological Polyakov loop potentials used in the literature and examine their correlation with thermal quarks in~a manner similar to the PNJL and PQM models.  
We tune our setup to be consistent with lattice results;
in particular, we adjust our model parameters to reproduce a pion mass of $m_\pi$$\sim$$640$~MeV 
and introduce a finite diquark source term, $\Delta_{\rm ex} = 45$~MeV, which is commonly used to stabilize lattice simulations.
The renormalization of the Polyakov loop is also taken into~account.

After calibrating our model to reproduce lattice results at $\mu_q = 0$, we study the Polyakov loop at finite $\mu_q$ and find reasonable agreement with the lattice data.  
Achieving this agreement requires a diquark gap of $\Delta$$\sim$$200$--$300$~MeV and a finite $\Delta_{\rm ex}$.  
If $\Delta$ is artificially set to zero, the~Polyakov loop reaches the deconfined solution too rapidly due to the large abundance of gapless~quarks.

We next delineate in-medium gluon propagators in the Landau--DeWitt gauge.  
At the one-loop level, it is known that at zero temperature, the color-singlet diquark gap protects gluons from screening by quarks.  
Thermal quarks overcoming the gap energy contribute to the screening as in the normal phase; 
hence, the~thermal modifications of the gluon mass can be used to study the nature of thermal excitations and how confinement affects~them.  

The Polyakov loop effects quenches the colored quarks and suppresses their contributions to the screening.  
These confining effects are important at the lowest temperature measured on the lattice, $T$$\sim$$44$~MeV, 
and should be taken into account when confronting theoretical computations with the lattice data.  
The thermal evolution of the screening masses at various $\mu_q$ appears reasonably consistent with the lattice results for the range of our parameters, 
$\alpha_s^q = 0.5$--2.0 and $m_g = 400$--500~MeV.

Overall, our analyses with the current lattice data suggest that thermal colored excitations in dense two-color QCD matter are strongly suppressed.
In the $\Phi = 1$ case, the~electric screening grows too rapidly compared to lattice results,
unless $\Delta$ is considerably larger than $\sim$$200$~MeV (this is unlikely however),
allowing thermal quarks to be effectively neglected even without Polyakov loop suppression.
Including the Polyakov loop increases the minimum excitation from $\sim$$\Delta$ to $\sim$$2\Delta$,
corresponding to the energy scale of baryonic (diquark) excitations.

To draw more definitive conclusions about confinement and deconfinement in dense QC$_2$D matter, however,
lattice data with smaller statistical fluctuations and better-controlled finite-volume and lattice-spacing effects are called for.
On the theoretical side, the~analytic computations should be improved by including higher-order loop corrections and genuinely non-perturbative~effects.

While this work focuses on the proof of principle about confinement/deconfinement in dense quark matter,
ultimately, the framework should be applied to three-color QCD. 
The quark back reaction to the gluon sector highly depends on the phase structure.
For instance, in~the case of the color-flavor-locked phase, 
there are both Debye and Meissner masses with which both electric and magnetic gluons are screened.
Another probability, as~suggested in the quarkyonic mater scenario, is that
the Fermi surface is dominated by color-singlet composites 
which do not strongly screen gluons, as~in the color-singlet diquarks in QC$_2$D.
These physics, dependent on the phase structure, are left for future~studies.

%

\vspace{6pt} 





\authorcontributions{Conceptualization, Y.K., T.K., and D.S.; formal analysis, Y.K. and T.K.; writing---original draft preparation, Y.K. and T.K.; writing---review and editing, T.K. and D.S.  All authors have read and agreed to the published version of the~manuscript.}

\funding{T.K. 
 was supported 
by JSPS KAKENHI Grant No. 23K03377;
D.S. by JSPS KAKENHI Grant No. 23H05439, No. 23H05439, and No.~25K17386.
}

\dataavailability{The data will be provided upon request. 
}

\acknowledgments{We thank A. Maas 
 and J. Skullerud for kindly providing us with their lattice data in Refs.~\cite{Boz:2019enj,Boz:2018crd},
and Itou for discussions.
}

\conflictsofinterest{The authors declare no conflicts of~interest.} 



\abbreviations{Abbreviations}{
The following abbreviations are used in this manuscript:\\

\noindent 
\begin{tabular}{@{}ll}
QC$_2$D & two-color quantum chromodynamics \\
QCD$_I$ & isospin QCD \\
EOS & equations of state \\
pYM & pure Yang--Mills \\
mYM & massive Yang--Mills \\
PNJL & Polyakov--Nambu--Lasinio \\
BFH & Brauner--Fukushima--Hidaka \\
BCS 
 & Bardeen--Cooper--Schrieffer \\
BEC & Bose--Einstein condensation \\

\end{tabular}
}
%

\begin{adjustwidth}{-\extralength}{0cm}

\reftitle{References}



\begin{thebibliography}{999}

\end{thebibliography}


\begin{thebibliography}{999}

\bibitem[Kogut et~al.(1999)Kogut, Stephanov, and Toublan]{Kogut:1999iv}
Kogut, J.B.; Stephanov, M.A.; Toublan, D.
\newblock {On two color QCD with baryon chemical potential}.
\newblock {\em Phys. Lett. B} {\bf 1999}, {\em 464},~183--191. 
\newblock {\url{https://doi.org/10.1016/S0370-2693(99)00971-5}}.

\bibitem[Kogut et~al.(2000)Kogut, Stephanov, Toublan, Verbaarschot, and
  Zhitnitsky]{Kogut:2000ek}
Kogut, J.B.; Stephanov, M.A.; Toublan, D.; Verbaarschot, J.J.M.; Zhitnitsky, A.
\newblock {QCD-like theories at finite baryon density}.
\newblock {\em Nucl. Phys. B} {\bf 2000}, {\em 582},~477--513.
\newblock {\url{https://doi.org/10.1016/S0550-3213(00)00242-X}}.

\bibitem[Suenaga et~al.(2023{\natexlab{a}})Suenaga, Murakami, Itou, and
  Iida]{Suenaga:2022uqn}
Suenaga, D.; Murakami, K.; Itou, E.; Iida, K.
\newblock {Probing the hadron mass spectrum in dense two-color QCD with the
  linear sigma model}.
\newblock {\em Phys. Rev. D} {\bf 2023}, {\em 107},~054001.
\newblock {\url{https://doi.org/10.1103/PhysRevD.107.054001}}.

\bibitem[Suenaga et~al.(2023{\natexlab{b}})Suenaga, Murakami, Itou, and
  Iida]{Suenaga:2023xwa}
Suenaga, D.; Murakami, K.; Itou, E.; Iida, K.
\newblock {Mass spectrum of spin-one hadrons in dense two-color QCD: Novel
  predictions by extended linear sigma model}. \emph{Phys. Rev. D} 
 {\bf 2023}, \emph{109}, 074031. 
\url{https://doi.org/10.1103/PhysRevD.109.074031}.

\bibitem[Kawaguchi and Suenaga(2024)]{Kawaguchi:2024iaw}
Kawaguchi, M.; Suenaga, D.
\newblock {Sound velocity peak induced by the chiral partner in dense two-color
  QCD}. \emph{Phys. Rev. D} {\bf 2024}, \emph{109}, 096034. \url{https://doi.org/10.1103/PhysRevD.109.096034}.

\bibitem[Suenaga(2025)]{Suenaga:2025sln}
Suenaga, D.
\newblock {Chiral Effective Model of Cold and Dense Two-Color QCD: The Linear
  Sigma Model Approach}.
\newblock {\em Symmetry} {\bf 2025}, {\em 17},~124.
\newblock {\url{https://doi.org/10.3390/sym17010124}}.

\bibitem[Sun et~al.(2007)Sun, He, and Zhuang]{Sun:2007fc}
Sun, G.f.; He, L.; Zhuang, P.
\newblock {BEC-BCS crossover in the Nambu-Jona-Lasinio model of QCD}.
\newblock {\em Phys. Rev. D} {\bf 2007}, {\em 75},~096004.
\newblock {\url{https://doi.org/10.1103/PhysRevD.75.096004}}.

\bibitem[Brauner et~al.(2009)Brauner, Fukushima, and Hidaka]{Brauner:2009gu}
Brauner, T.; Fukushima, K.; Hidaka, Y.
\newblock {Two-color quark matter: U(1)(A) restoration, superfluidity, and
  quarkyonic phase}.
\newblock {\em Phys. Rev. D} {\bf 2009}, {\em 80},~074035;
\newblock Erratum in: \emph{Phys. Rev. D} \textbf{2010}, \emph{81}, 119904.
  {\url{https://doi.org/10.1103/PhysRevD.81.119904}}.

\bibitem[Strodthoff and von Smekal(2014)]{Strodthoff:2013cua}
Strodthoff, N.; von Smekal, L.
\newblock {Polyakov-Quark-Meson-Diquark Model for two-color QCD}.
\newblock {\em Phys. Lett. B} {\bf 2014}, {\em 731},~350--357.
\newblock {\url{https://doi.org/10.1016/j.physletb.2014.03.008}}.

\bibitem[Strodthoff et~al.(2012)Strodthoff, Schaefer, and von
  Smekal]{Strodthoff:2011tz}
Strodthoff, N.; Schaefer, B.J.; von Smekal, L.
\newblock {Quark-meson-diquark model for two-color QCD}.
\newblock {\em Phys. Rev. D} {\bf 2012}, {\em 85},~074007.
\newblock {\url{https://doi.org/10.1103/PhysRevD.85.074007}}.

\bibitem[Itou(2025)]{Itou:2025vcy}
Itou, E.
\newblock {Lattice results for the equation of state in dense QCD-like
  theories.} \emph{arXiv} {\bf 2025}. \url{https://doi.org/10.48550/arXiv.2508.03090}.

\bibitem[Iida et~al.(2024)Iida, Itou, Murakami, and Suenaga]{Iida:2024irv}
Iida, K.; Itou, E.; Murakami, K.; Suenaga, D.
\newblock {Lattice study on finite density QC$_2$D towards zero temperature.} \emph{arXiv}
  {\bf 2024}. \url{https://doi.org/10.48550/arXiv.2405.20566}.

\bibitem[Iida et~al.(2020)Iida, Itou, and Lee]{Iida:2019rah}
Iida, K.; Itou, E.; Lee, T.G.
\newblock {Two-colour QCD phases and the topology at low temperature and high
  density}.
\newblock {\em J. High Energy Phys.} {\bf 2020}, {\em 2020},~181.
\newblock {\url{https://doi.org/10.1007/JHEP01(2020)181}}.

\bibitem[Iida et~al.(2021)Iida, Itou, and Lee]{Iida:2020emi}
Iida, K.; Itou, E.; Lee, T.G.
\newblock {Relative scale setting for two-color QCD with $N_f$=2 Wilson
  fermions}.
\newblock {\em Prog. Theor. Exp. Phys.} {\bf 2021}, {\em 2021},~013B05.
\newblock {\url{https://doi.org/10.1093/ptep/ptaa170}}.

\bibitem[Boz et~al.(2020)Boz, Giudice, Hands, and Skullerud]{Boz:2019enj}
Boz, T.; Giudice, P.; Hands, S.; Skullerud, J.I.
\newblock {Dense two-color QCD towards continuum and chiral limits}.
\newblock {\em Phys. Rev. D} {\bf 2020}, {\em 101},~074506.
\newblock {\url{https://doi.org/10.1103/PhysRevD.101.074506}}.

\bibitem[Boz et~al.(2013)Boz, Cotter, Fister, Mehta, and
  Skullerud]{Boz:2013rca}
Boz, T.; Cotter, S.; Fister, L.; Mehta, D.; Skullerud, J.I.
\newblock {Phase transitions and gluodynamics in 2-colour matter at high
  density}.
\newblock {\em Eur. Phys. J. A} {\bf 2013}, {\em 49},~87.
\newblock {\url{https://doi.org/10.1140/epja/i2013-13087-6}}.

\bibitem[Cotter et~al.(2013)Cotter, Giudice, Hands, and
  Skullerud]{Cotter:2012mb}
Cotter, S.; Giudice, P.; Hands, S.; Skullerud, J.I.
\newblock {Towards the phase diagram of dense two-color matter}.
\newblock {\em Phys. Rev. D} {\bf 2013}, {\em 87},~034507.
\newblock {\url{https://doi.org/10.1103/PhysRevD.87.034507}}.

\bibitem[Hands et~al.(2011)Hands, Kenny, Kim, and Skullerud]{Hands:2011ye}
Hands, S.; Kenny, P.; Kim, S.; Skullerud, J.I.
\newblock {Lattice Study of Dense Matter with Two Colors and Four Flavors}.
\newblock {\em Eur. Phys. J. A} {\bf 2011}, {\em 47},~60.
\newblock {\url{https://doi.org/10.1140/epja/i2011-11060-1}}.

\bibitem[Astrakhantsev et~al.(2020)Astrakhantsev, Braguta, Ilgenfritz, Kotov,
  and Nikolaev]{Astrakhantsev:2020tdl}
Astrakhantsev, N.; Braguta, V.; Ilgenfritz, E.; Kotov, A.; Nikolaev, A.
\newblock {Lattice study of thermodynamic properties of dense QC$_2$D}.
\newblock {\em Phys. Rev. D} {\bf 2020}, {\em 102},~074507.
\newblock {\url{https://doi.org/10.1103/PhysRevD.102.074507}}.

\bibitem[Bornyakov et~al.(2020)Bornyakov, Braguta, Nikolaev, and
  Rogalyov]{Bornyakov:2020kyz}
Bornyakov, V.G.; Braguta, V.V.; Nikolaev, A.A.; Rogalyov, R.N.
\newblock {Effects of Dense Quark Matter on Gluon Propagators in Lattice
  QC$_2$D}.
\newblock {\em Phys. Rev. D} {\bf 2020}, {\em 102},~114511.
\newblock {\url{https://doi.org/10.1103/PhysRevD.102.114511}}.

\bibitem[Muroya et~al.(2003)Muroya, Nakamura, and Nonaka]{Muroya:2002ry}
Muroya, S.; Nakamura, A.; Nonaka, C.
\newblock {Behavior of hadrons at finite density: Lattice study of color SU(2)
  QCD}.
\newblock {\em Phys. Lett. B} {\bf 2003}, {\em 551},~305--310.
\newblock {\url{https://doi.org/10.1016/S0370-2693(02)03065-4}}.

\bibitem[Abbott et~al.(2023)Abbott, Detmold, Romero-L{\'o}pez, Davoudi, Illa,
  Parre{\~n}o, Perry, Shanahan, and Wagman]{Abbott:2023coj}
\textls[-10]{Abbott, R.; Detmold, W.; Romero-L{\'o}pez, F.; Davoudi, Z.; Illa, M.;
  Parre{\~n}o, A.; Perry, R.J.; Shanahan, P.E.; Wagman, M.L.
\newblock {Lattice quantum chromodynamics at large isospin density}.
\newblock {\em Phys. Rev. D} {\bf 2023}, {\em 108},~114506.
\newblock {\url{https://doi.org/10.1103/PhysRevD.108.114506}}.}

\bibitem[Abbott et~al.(2025)Abbott, Detmold, Illa, Parre{\~n}o, Perry,
  Romero-L{\'o}pez, Shanahan, and Wagman]{Abbott:2024vhj}
Abbott, R.; Detmold, W.; Illa, M.; Parre{\~n}o, A.; Perry, R.J.;
  Romero-L{\'o}pez, F.; Shanahan, P.E.; Wagman, M.L.
\newblock {QCD Constraints on Isospin-Dense Matter and the Nuclear Equation of
  State}.
\newblock {\em Phys. Rev. Lett.} {\bf 2025}, {\em 134},~011903.
\newblock {\url{https://doi.org/10.1103/PhysRevLett.134.011903}}.

\bibitem[Brandt et~al.(2023)Brandt, Cuteri, and Endrodi]{Brandt:2022hwy}
Brandt, B.B.; Cuteri, F.; Endrodi, G.
\newblock {Equation of state and speed of sound of isospin-asymmetric QCD on
  the lattice}.
\newblock {\em J. High Energy Phys.} {\bf 2023}, {\em 7},~55.
\newblock {\url{https://doi.org/10.1007/JHEP07(2023)055}}.

\bibitem[Brandt et~al.(2018)Brandt, Endrodi, and Schmalzbauer]{Brandt:2017oyy}
Brandt, B.B.; Endrodi, G.; Schmalzbauer, S.
\newblock {QCD phase diagram for nonzero isospin-asymmetry}.
\newblock {\em Phys. Rev. D} {\bf 2018}, {\em 97},~054514.
\newblock {\url{https://doi.org/10.1103/PhysRevD.97.054514}}.

\bibitem[Fujimoto(2024{\natexlab{a}})]{Fujimoto:2024pcd}
Fujimoto, Y.
\newblock {Interplay between the weak-coupling results and the lattice data in
  dense QCD.} \emph{arXiv} {\bf 2024}. \url{https://doi.org/10.48550/arXiv.2408.12514}.

\bibitem[Fujimoto(2024{\natexlab{b}})]{Fujimoto:2023mvc}
Fujimoto, Y.
\newblock {Enhanced contribution of the pairing gap to the QCD equation of
  state at large isospin chemical potential}.
\newblock {\em Phys. Rev. D} {\bf 2024}, {\em 109},~054035.
\newblock {\url{https://doi.org/10.1103/PhysRevD.109.054035}}.

\bibitem[Fukushima and Minato(2025)]{Fukushima:2024gmp}
Fukushima, K.; Minato, S.
\newblock {Speed of sound and trace anomaly in a unified treatment of the
  two-color diquark superfluid, the pion-condensed high-isospin matter, and the
  2SC quark matter}.
\newblock {\em Phys. Rev. D} {\bf 2025}, {\em 111},~094006.
\newblock {\url{https://doi.org/10.1103/PhysRevD.111.094006}}.

\bibitem[Kojo et~al.(2015)Kojo, Powell, Song, and Baym]{Kojo:2014rca}
Kojo, T.; Powell, P.D.; Song, Y.; Baym, G.
\newblock {Phenomenological QCD equation of state for massive neutron stars}.
\newblock {\em Phys. Rev. D} {\bf 2015}, {\em 91},~045003.
\newblock {\url{https://doi.org/10.1103/PhysRevD.91.045003}}.

\bibitem[Kojo et~al.(2022)Kojo, Baym, and Hatsuda]{Kojo:2021wax}
Kojo, T.; Baym, G.; Hatsuda, T.
\newblock {Implications of NICER for Neutron Star Matter: The QHC21 Equation of
  State}.
\newblock {\em Astrophys. J.} {\bf 2022}, {\em 934},~46.
\newblock {\url{https://doi.org/10.3847/1538-4357/ac7876}}.

\bibitem[Gei{\ss}el et~al.(2025)Gei{\ss}el, Gorda, and Braun]{Geissel:2025vnp}
Gei{\ss}el, A.; Gorda, T.; Braun, J.
\newblock {Color superconductivity under neutron-star conditions at
  next-to-leading order.} \emph{arXiv} {\bf 2025}. \url{https://doi.org/10.48550/arXiv.2504.03834}.

\bibitem[Gei{\ss}el et~al.(2024)Gei{\ss}el, Gorda, and Braun]{Geissel:2024nmx}
Gei{\ss}el, A.; Gorda, T.; Braun, J.
\newblock {Pressure and speed of sound in two-flavor color-superconducting
  quark matter at next-to-leading order}.
\newblock {\em Phys. Rev. D} {\bf 2024}, {\em 110},~014034.
\newblock {\url{https://doi.org/10.1103/PhysRevD.110.014034}}.

\bibitem[Kojo and Suenaga(2022)]{Kojo:2021hqh}
Kojo, T.; Suenaga, D.
\newblock {Peaks of sound velocity in two color dense QCD: Quark saturation
  effects and semishort range correlations}.
\newblock {\em Phys. Rev. D} {\bf 2022}, {\em 105},~076001.
\newblock {\url{https://doi.org/10.1103/PhysRevD.105.076001}}.

\bibitem[Chiba and Kojo(2024)]{Chiba:2023ftg}
Chiba, R.; Kojo, T.
\newblock {Sound velocity peak and conformality in isospin QCD}.
\newblock {\em Phys. Rev. D} {\bf 2024}, {\em 109},~076006.
\newblock {\url{https://doi.org/10.1103/PhysRevD.109.076006}}.

\bibitem[Kojo(2025)]{Kojo:2025vcq}
Kojo, T.
\newblock {Stiffening of matter in quark{\textendash}hadron continuity: A
  mini-review}.
\newblock {\em J. Subat. Part. Cosmol.} {\bf 2025}, {\em 4},~100088.
\newblock {\url{https://doi.org/10.1016/j.jspc.2025.100088}}.

\bibitem[von Smekal(2012)]{vonSmekal:2012vx}
von Smekal, L.
\newblock {Universal Aspects of QCD-like Theories}.
\newblock {\em Nucl. Phys. B Proc. Suppl.} {\bf 2012}, {\em 228},~179--220.
\newblock {\url{https://doi.org/10.1016/j.nuclphysbps.2012.06.006}}.

\bibitem[Kamikado et~al.(2013)Kamikado, Strodthoff, von Smekal, and
  Wambach]{Kamikado:2012bt}
Kamikado, K.; Strodthoff, N.; von Smekal, L.; Wambach, J.
\newblock {Fluctuations in the quark-meson model for QCD with isospin chemical
  potential}.
\newblock {\em Phys. Lett. B} {\bf 2013}, {\em 718},~1044--1053.
\newblock {\url{https://doi.org/10.1016/j.physletb.2012.11.055}}.

\bibitem[Andersen et~al.(2019)Andersen, Adhikari, and
  Kneschke]{Andersen:2018qkq}
Andersen, J.O.; Adhikari, P.; Kneschke, P.
\newblock {Pion condensation and QCD phase diagram at finite isospin density}. In Proceedings of the XIII Quark Confinement and the Hadron Spectrum---Confinement 2018, Maynooth, Ireland, 31 July--6 August 2018; 
197. 
\newblock {\url{https://doi.org/10.22323/1.336.0197}}.

\bibitem[Chiba et~al.(2024)Chiba, Kojo, and Suenaga]{Chiba:2024cny}
Chiba, R.; Kojo, T.; Suenaga, D.
\newblock {Thermal effects on sound velocity peak and conformality in isospin
  QCD.} \emph{Phys. Rev. D} {\bf 2024}, \mbox{\emph{110}, 054037.} \url{https://doi.org/10.1103/PhysRevD.110.054037}.

\bibitem[Brandt et~al.(2025)Brandt, Chelnokov, Endrodi, Marko, Scheid, and von
  Smekal]{Brandt:2025tkg}
Brandt, B.B.; Chelnokov, V.; Endrodi, G.; Marko, G.; Scheid, D.; von Smekal, L.
\newblock {Renormalization group invariant mean-field model for QCD at finite
  isospin density.} \emph{Phys. Rev. D} {\bf 2025}, \emph{112}, 054038. \url{https://doi.org/10.1103/fryz-f3vw}.

\bibitem[Carlomagno et~al.(2024)Carlomagno, Gomez~Dumm, and
  Scoccola]{Carlomagno:2024xmi}
Carlomagno, J.P.; Gomez~Dumm, D.; Scoccola, N.N.
\newblock {Cold isospin asymmetric baryonic rich matter in nonlocal NJL-like
  models}.
\newblock {\em Phys. Rev. D} {\bf 2024}, {\em 109},~094041.
\newblock {\url{https://doi.org/10.1103/PhysRevD.109.094041}}.

\bibitem[Lopes et~al.(2025)Lopes, Duarte, Farias, and O.~Ramos]{Lopes:2025rvn}
Lopes, B.S.; Duarte, D.C.; Farias, R.L.S.; Ramos, R.O.
\newblock {Speed of sound peak in isospin QCD: A natural prediction of the
  Medium Separation Scheme.} \emph{arXiv} {\bf 2025}. \url{https://doi.org/10.48550/arXiv.2507.14343}.

\bibitem[Ayala et~al.(2025)Ayala, Lopes, Farias, and Parra]{Ayala:2024sqm}
Ayala, A.; Lopes, B.S.; Farias, R.L.S.; Parra, L.C.
\newblock {On the origin of the peak of the sound velocity for isospin
  imbalanced strongly interacting matter}.
\newblock {\em Phys. Lett. B} {\bf 2025}, {\em 864},~139396.
\newblock {\url{https://doi.org/10.1016/j.physletb.2025.139396}}.

\bibitem[Masuda et~al.(2013{\natexlab{a}})Masuda, Hatsuda, and
  Takatsuka]{Masuda:2012kf}
Masuda, K.; Hatsuda, T.; Takatsuka, T.
\newblock {Hadron-Quark Crossover and Massive Hybrid Stars with Strangeness}.
\newblock {\em Astrophys. J.} {\bf 2013}, {\em 764},~12.
\newblock {\url{https://doi.org/10.1088/0004-637X/764/1/12}}.

\bibitem[Masuda et~al.(2013{\natexlab{b}})Masuda, Hatsuda, and
  Takatsuka]{Masuda:2012ed}
Masuda, K.; Hatsuda, T.; Takatsuka, T.
\newblock {Hadron--quark crossover and massive hybrid stars}.
\newblock {\em Prog. Theor. Exp. Phys.} {\bf 2013}, {\em 2013},~073D01.
\newblock {\url{https://doi.org/10.1093/ptep/ptt045}}.

\bibitem[Kojo(2016)]{Kojo:2015fua}
Kojo, T.
\newblock {Phenomenological neutron star equations of state: 3-window modeling
  of QCD matter}.
\newblock {\em Eur. Phys. J. A} {\bf 2016}, {\em 52},~51.
\newblock {\url{https://doi.org/10.1140/epja/i2016-16051-0}}.

\bibitem[Baym et~al.(2019)Baym, Furusawa, Hatsuda, Kojo, and
  Togashi]{Baym:2019iky}
Baym, G.; Furusawa, S.; Hatsuda, T.; Kojo, T.; Togashi, H.
\newblock {New Neutron Star Equation of State with Quark-Hadron Crossover}.
\newblock {\em Astrophys. J.} {\bf 2019}, {\em 885},~42.
\newblock {\url{https://doi.org/10.3847/1538-4357/ab441e}}.

\bibitem[Ayriyan et~al.(2021)Ayriyan, Blaschke, Grunfeld, Alvarez-Castillo,
  Grigorian, and Abgaryan]{Ayriyan:2021prr}
Ayriyan, A.; Blaschke, D.; Grunfeld, A.G.; Alvarez-Castillo, D.; Grigorian, H.;
  Abgaryan, V.
\newblock {Bayesian analysis of multimessenger M-R data with interpolated
  hybrid EoS.} \emph{Eur. Phys. J. A} {\bf 2021}, \emph{57}, 318. \url{https://doi.org/10.1140/epja/s10050-021-00619-0}.

\bibitem[Minamikawa et~al.(2021)Minamikawa, Kojo, and
  Harada]{Minamikawa:2020jfj}
Minamikawa, T.; Kojo, T.; Harada, M.
\newblock {Quark-hadron crossover equations of state for neutron stars:
  constraining the chiral invariant mass in a parity doublet model}.
\newblock {\em Phys. Rev. C} {\bf 2021}, {\em 103},~045205.
\newblock {\url{https://doi.org/10.1103/PhysRevC.103.045205}}.

\bibitem[Fraga et~al.(2014)Fraga, Kurkela, and Vuorinen]{Fraga:2013qra}
Fraga, E.S.; Kurkela, A.; Vuorinen, A.
\newblock {Interacting quark matter equation of state for compact stars}.
\newblock {\em Astrophys. J. Lett.} {\bf 2014}, {\em 781},~L25.
\newblock {\url{https://doi.org/10.1088/2041-8205/781/2/L25}}.

\bibitem[Kurkela et~al.(2014)Kurkela, Fraga, Schaffner-Bielich, and
  Vuorinen]{Kurkela:2014vha}
Kurkela, A.; Fraga, E.S.; Schaffner-Bielich, J.; Vuorinen, A.
\newblock {Constraining neutron star matter with Quantum Chromodynamics}.
\newblock {\em Astrophys. J.} {\bf 2014}, {\em 789},~127.
\newblock {\url{https://doi.org/10.1088/0004-637X/789/2/127}}.

\bibitem[Komoltsev and Kurkela(2022)]{Komoltsev:2021jzg}
Komoltsev, O.; Kurkela, A.
\newblock {How Perturbative QCD Constrains the Equation of State at
  Neutron-Star Densities}.
\newblock {\em Phys. Rev. Lett.} {\bf 2022}, {\em 128},~202701.
\newblock {\url{https://doi.org/10.1103/PhysRevLett.128.202701}}.

\bibitem[Gorda et~al.(2023)Gorda, Komoltsev, and Kurkela]{Gorda:2022jvk}
Gorda, T.; Komoltsev, O.; Kurkela, A.
\newblock {Ab-initio QCD Calculations Impact the Inference of the
  Neutron-star-matter Equation of State}.
\newblock {\em Astrophys. J.} {\bf 2023}, {\em 950},~107.
\newblock {\url{https://doi.org/10.3847/1538-4357/acce3a}}.

\bibitem[Kojo(2021)]{Kojo:2021ugu}
Kojo, T.
\newblock {Stiffening of matter in quark-hadron continuity}.
\newblock {\em Phys. Rev. D} {\bf 2021}, {\em 104},~074005.
\newblock {\url{https://doi.org/10.1103/PhysRevD.104.074005}}.

\bibitem[Fujimoto et~al.(2024)Fujimoto, Kojo, and McLerran]{Fujimoto:2023mzy}
Fujimoto, Y.; Kojo, T.; McLerran, L.D.
\newblock {Momentum Shell in Quarkyonic Matter from Explicit Duality: A Dual
  Model for Cold, Dense QCD}.
\newblock {\em Phys. Rev. Lett.} {\bf 2024}, {\em 132},~112701.
\newblock {\url{https://doi.org/10.1103/PhysRevLett.132.112701}}.

\bibitem[Tajima et~al.(2025)Tajima, Iida, Kojo, and Liang]{Tajima:2024qzj}
Tajima, H.; Iida, K.; Kojo, T.; Liang, H.
\newblock {Tripling Fluctuations and Peaked Sound Speed in Fermionic Matter}.
\newblock {\em Phys. Rev. Lett.} {\bf 2025}, {\em 135},~042701.
\newblock {\url{https://doi.org/10.1103/4ywp-752m}}.

\bibitem[Fujimoto et~al.(2024)Fujimoto, Kojo, and McLerran]{Fujimoto:2024doc}
Fujimoto, Y.; Kojo, T.; McLerran, L.
\newblock {Quarkyonic matter pieces together the hyperon puzzle.} \emph{arXiv} {\bf 2024}. \url{https://doi.org/10.48550/arXiv.2410.22758}.

\bibitem[Ivanytskyi(2025)]{Ivanytskyi:2025cnn}
Ivanytskyi, O.
\newblock {Quarkyonic picture of isospin QCD}.
\newblock {\em Phys. Rev. D} {\bf 2025}, {\em 112},~034001.
\newblock {\url{https://doi.org/10.1103/831v-8mp4}}.

\bibitem[Kojo and Baym(2014)]{Kojo:2014vja}
Kojo, T.; Baym, G.
\newblock {Color screening in cold quark matter}.
\newblock {\em Phys. Rev. D} {\bf 2014}, {\em 89},~125008.
\newblock {\url{https://doi.org/10.1103/PhysRevD.89.125008}}.

\bibitem[Suenaga and Kojo(2019)]{Suenaga:2019jjv}
Suenaga, D.; Kojo, T.
\newblock {Gluon propagator in two-color dense QCD: Massive Yang-Mills approach
  at one-loop}.
\newblock {\em Phys. Rev. D} {\bf 2019}, {\em 100},~076017.
\newblock {\url{https://doi.org/10.1103/PhysRevD.100.076017}}.

\bibitem[Kojo and Suenaga(2021)]{Kojo:2021knn}
Kojo, T.; Suenaga, D.
\newblock {Thermal quarks and gluon propagators in two-color dense QCD}.
\newblock {\em Phys. Rev. D} {\bf 2021}, {\em 103},~094008.
\newblock {\url{https://doi.org/10.1103/PhysRevD.103.094008}}.

\bibitem[Contant and Huber(2020)]{Contant:2019lwf}
Contant, R.; Huber, M.Q.
\newblock {Dense two-color QCD from Dyson-Schwinger equations}.
\newblock {\em Phys. Rev. D} {\bf 2020}, {\em 101},~014016.
\newblock {\url{https://doi.org/10.1103/PhysRevD.101.014016}}.

\bibitem[Contant and Huber(2017)]{Contant:2017gtz}
Contant, R.; Huber, M.Q.
\newblock {Phase structure and propagators at nonvanishing temperature for QCD
  and QCD-like theories}.
\newblock {\em Phys. Rev. D} {\bf 2017}, {\em 96},~074002.
\newblock {\url{https://doi.org/10.1103/PhysRevD.96.074002}}.

\bibitem[Rischke et~al.(2001)Rischke, Son, and Stephanov]{Rischke:2000cn}
Rischke, D.H.; Son, D.T.; Stephanov, M.A.
\newblock {Asymptotic deconfinement in high density QCD}.
\newblock {\em Phys. Rev. Lett.} {\bf 2001}, {\em 87},~062001.
\newblock {\url{https://doi.org/10.1103/PhysRevLett.87.062001}}.

\bibitem[Cohen and Sen(2015)]{Cohen:2015soa}
Cohen, T.D.; Sen, S.
\newblock {Deconfinement Transition at High Isospin Chemical Potential and Low
  Temperature}.
\newblock {\em Nucl. Phys. A} {\bf 2015}, {\em 942},~39--53.
\newblock {\url{https://doi.org/10.1016/j.nuclphysa.2015.07.018}}.

\bibitem[McLerran and Pisarski(2007)]{McLerran:2007qj}
McLerran, L.; Pisarski, R.D.
\newblock {Phases of cold, dense quarks at large N(c)}.
\newblock {\em Nucl. Phys. A} {\bf 2007}, {\em 796},~83--100.
\newblock {\url{https://doi.org/10.1016/j.nuclphysa.2007.08.013}}.

\bibitem[Polyakov(1978)]{Polyakov:1978vu}
Polyakov, A.M.
\newblock {Thermal Properties of Gauge Fields and Quark Liberation}.
\newblock {\em Phys. Lett. B} {\bf 1978}, {\em 72},~477--480.
\newblock {\url{https://doi.org/10.1016/0370-2693(78)90737-2}}.

\bibitem[Fukushima and Skokov(2017)]{Fukushima:2017csk}
Fukushima, K.; Skokov, V.
\newblock {Polyakov loop modeling for hot QCD}.
\newblock {\em Prog. Part. Nucl. Phys.} {\bf 2017}, {\em 96},~154--199.
\newblock {\url{https://doi.org/10.1016/j.ppnp.2017.05.002}}.

\bibitem[Braun et~al.(2010)Braun, Gies, and Pawlowski]{Braun:2007bx}
Braun, J.; Gies, H.; Pawlowski, J.M.
\newblock {Quark Confinement from Color Confinement}.
\newblock {\em Phys. Lett. B} {\bf 2010}, {\em 684},~262--267.
\newblock {\url{https://doi.org/10.1016/j.physletb.2010.01.009}}.

\bibitem[Fukushima and Kashiwa(2013)]{Fukushima:2012qa}
Fukushima, K.; Kashiwa, K.
\newblock {Polyakov loop and QCD thermodynamics from the gluon and ghost
  propagators}.
\newblock {\em Phys. Lett. B} {\bf 2013}, {\em 723},~360--364.
\newblock {\url{https://doi.org/10.1016/j.physletb.2013.05.037}}.

\bibitem[Fukushima and Su(2013)]{Fukushima:2013xsa}
Fukushima, K.; Su, N.
\newblock {Stabilizing perturbative Yang-Mills thermodynamics with Gribov
  quantization}.
\newblock {\em Phys. Rev. D} {\bf 2013}, {\em 88},~076008.
\newblock {\url{https://doi.org/10.1103/PhysRevD.88.076008}}.

\bibitem[Reinosa et~al.(2015{\natexlab{a}})Reinosa, Serreau, Tissier, and
  Wschebor]{Reinosa:2014ooa}
Reinosa, U.; Serreau, J.; Tissier, M.; Wschebor, N.
\newblock {Deconfinement transition in SU($N$) theories from perturbation
  theory}.
\newblock {\em Phys. Lett. B} {\bf 2015}, {\em 742},~61--68.
\newblock {\url{https://doi.org/10.1016/j.physletb.2015.01.006}}.

\bibitem[Reinosa et~al.(2015{\natexlab{b}})Reinosa, Serreau, Tissier, and
  Wschebor]{Reinosa:2014zta}
Reinosa, U.; Serreau, J.; Tissier, M.; Wschebor, N.
\newblock {Deconfinement transition in SU(2) Yang-Mills theory: A two-loop
  study}.
\newblock {\em Phys. Rev. D} {\bf 2015}, {\em 91},~045035.
\newblock {\url{https://doi.org/10.1103/PhysRevD.91.045035}}.

\bibitem[Boz et~al.(2019)Boz, Hajizadeh, Maas, and Skullerud]{Boz:2018crd}
Boz, T.; Hajizadeh, O.; Maas, A.; Skullerud, J.I.
\newblock {Finite-density gauge correlation functions in QC2D}.
\newblock {\em Phys. Rev. D} {\bf 2019}, {\em 99},~074514.
\newblock {\url{https://doi.org/10.1103/PhysRevD.99.074514}}.

\bibitem[Tissier and Wschebor(2011)]{Tissier:2011ey}
Tissier, M.; Wschebor, N.
\newblock {An Infrared Safe perturbative approach to Yang-Mills correlators}.
\newblock {\em Phys. Rev. D} {\bf 2011}, {\em 84},~045018.
\newblock {\url{https://doi.org/10.1103/PhysRevD.84.045018}}.

\bibitem[Tissier and Wschebor(2010)]{Tissier:2010ts}
Tissier, M.; Wschebor, N.
\newblock {Infrared propagators of Yang-Mills theory from perturbation theory}.
\newblock {\em Phys. Rev. D} {\bf 2010}, {\em 82},~101701.
\newblock {\url{https://doi.org/10.1103/PhysRevD.82.101701}}.

\bibitem[Reinosa et~al.(2017)Reinosa, Serreau, Tissier, and
  Wschebor]{Reinosa:2017qtf}
Reinosa, U.; Serreau, J.; Tissier, M.; Wschebor, N.
\newblock {How nonperturbative is the infrared regime of Landau gauge
  Yang-Mills correlators?}
\newblock {\em Phys. Rev. D} {\bf 2017}, {\em 96},~014005.
\newblock {\url{https://doi.org/10.1103/PhysRevD.96.014005}}.

\bibitem[Gracey et~al.(2019)Gracey, Pel{\'a}ez, Reinosa, and
  Tissier]{Gracey:2019xom}
Gracey, J.A.; Pel{\'a}ez, M.; Reinosa, U.; Tissier, M.
\newblock {Two loop calculation of Yang-Mills propagators in the Curci-Ferrari
  model}.
\newblock {\em Phys. Rev. D} {\bf 2019}, {\em 100},~034023.
\newblock {\url{https://doi.org/10.1103/PhysRevD.100.034023}}.

\bibitem[Pel{\'a}ez et~al.(2014)Pel{\'a}ez, Tissier, and
  Wschebor]{Pelaez:2014mxa}
Pel{\'a}ez, M.; Tissier, M.; Wschebor, N.
\newblock {Two-point correlation functions of QCD in the Landau gauge}.
\newblock {\em Phys. Rev. D} {\bf 2014}, {\em 90},~065031.
\newblock {\url{https://doi.org/10.1103/PhysRevD.90.065031}}.

\bibitem[Pel{\'a}ez et~al.(2017)Pel{\'a}ez, Reinosa, Serreau, Tissier, and
  Wschebor]{Pelaez:2017bhh}
Pel{\'a}ez, M.; Reinosa, U.; Serreau, J.; Tissier, M.; Wschebor, N.
\newblock {Small parameters in infrared quantum chromodynamics}.
\newblock {\em Phys. Rev. D} {\bf 2017}, {\em 96},~114011.
\newblock {\url{https://doi.org/10.1103/PhysRevD.96.114011}}.

\bibitem[Pel{\'a}ez et~al.(2021)Pel{\'a}ez, Reinosa, Serreau, Tissier, and
  Wschebor]{Pelaez:2021tpq}
Pel{\'a}ez, M.; Reinosa, U.; Serreau, J.; Tissier, M.; Wschebor, N.
\newblock {A window on infrared QCD with small expansion parameters}.
\newblock {\em Rept. Prog. Phys.} {\bf 2021}, {\em 84},~124202.
\newblock {\url{https://doi.org/10.1088/1361-6633/ac36b8}}.

\bibitem[Gross et~al.(1981)Gross, Pisarski, and Yaffe]{Gross:1980br}
Gross, D.J.; Pisarski, R.D.; Yaffe, L.G.
\newblock {QCD and Instantons at Finite Temperature}.
\newblock {\em Rev. Mod. Phys.} {\bf 1981}, {\em 53},~43.
\newblock {\url{https://doi.org/10.1103/RevModPhys.53.43}}.

\bibitem[Weiss(1981)]{Weiss:1980rj}
Weiss, N.
\newblock {The Effective Potential for the Order Parameter of Gauge Theories at
  Finite Temperature}.
\newblock {\em Phys. Rev. D} {\bf 1981}, {\em 24},~475.
\newblock {\url{https://doi.org/10.1103/PhysRevD.24.475}}.

\bibitem[Gribov(1978)]{Gribov:1977wm}
Gribov, V.N.
\newblock {Quantization of Nonabelian Gauge Theories}.
\newblock {\em Nucl. Phys. B} {\bf 1978}, {\em 139},~1--19.
\newblock {\url{https://doi.org/10.1016/0550-3213(78)90175-X}}.

\bibitem[Vandersickel and Zwanziger(2012)]{Vandersickel:2012tz}
Vandersickel, N.; Zwanziger, D.
\newblock {The Gribov problem and QCD dynamics}.
\newblock {\em Phys. Rept.} {\bf 2012}, {\em 520},~175--251.
\newblock {\url{https://doi.org/10.1016/j.physrep.2012.07.003}}.

\bibitem[Canfora et~al.(2015)Canfora, Dudal, Justo, Pais, Rosa, and
  Vercauteren]{Canfora:2015yia}
Canfora, F.E.; Dudal, D.; Justo, I.F.; Pais, P.; Rosa, L.; Vercauteren, D.
\newblock {Effect of the Gribov horizon on the Polyakov loop and vice~versa}.
\newblock {\em Eur. Phys. J. C} {\bf 2015}, {\em 75},~326.
\newblock {\url{https://doi.org/10.1140/epjc/s10052-015-3546-y}}.

\bibitem[Mei et~al.(2025)Mei, Wang, and Huang]{Mei:2025kha}
Mei, J.; Wang, L.; Huang, M.
\newblock {Neural network extraction of chromo-electric and chromo-magnetic
  gluon masses.} \emph{arXiv} {\bf 2025}. \url{https://doi.org/10.48550/arXiv.2507.22012}.
%

\bibitem[Son and Stephanov(2001{\natexlab{a}})]{Son:2000xc}
Son, D.T.; Stephanov, M.A.
\newblock {QCD at finite isospin density}.
\newblock {\em Phys. Rev. Lett.} {\bf 2001}, {\em 86},~592--595.
\newblock {\url{https://doi.org/10.1103/PhysRevLett.86.592}}.

\bibitem[Son and Stephanov(2001{\natexlab{b}})]{Son:2000by}
Son, D.T.; Stephanov, M.A.
\newblock {QCD at finite isospin density: From pion to quark - anti-quark
  condensation}.
\newblock {\em Phys. Atom. Nucl.} {\bf 2001}, {\em 64},~834--842.
\newblock {\url{https://doi.org/10.1134/1.1378872}}.

\bibitem[Suenaga and Kojo(2021)]{Suenaga:2021bjz}
Suenaga, D.; Kojo, T.
\newblock {Delineating chiral separation effect in two-color dense QCD}.
\newblock {\em Phys. Rev. D} {\bf 2021}, {\em 104},~034038.
\newblock {\url{https://doi.org/10.1103/PhysRevD.104.034038}}.

\bibitem[Pisarski and Rischke(2000)]{Pisarski:1999bf}
Pisarski, R.D.; Rischke, D.H.
\newblock {Gaps and critical temperature for color superconductivity}.
\newblock {\em Phys. Rev. D} {\bf 2000}, {\em 61},~051501.
\newblock {\url{https://doi.org/10.1103/PhysRevD.61.051501}}.

\bibitem[Huang and Shovkovy(2004{\natexlab{a}})]{Huang:2004bg}
Huang, M.; Shovkovy, I.A.
\newblock {Chromomagnetic instability in dense quark matter}.
\newblock {\em Phys. Rev. D} {\bf 2004}, {\em 70},~051501.
\newblock {\url{https://doi.org/10.1103/PhysRevD.70.051501}}.

\bibitem[Huang and Shovkovy(2004{\natexlab{b}})]{Huang:2004am}
Huang, M.; Shovkovy, I.A.
\newblock {Screening masses in neutral two-flavor color superconductor}.
\newblock {\em Phys. Rev. D} {\bf 2004}, {\em 70},~094030.
\newblock {\url{https://doi.org/10.1103/PhysRevD.70.094030}}.

\bibitem[Shovkovy and Huang(2003)]{Shovkovy:2003uu}
Shovkovy, I.; Huang, M.
\newblock {Gapless two flavor color superconductor}.
\newblock {\em Phys. Lett. B} {\bf 2003}, {\em 564},~205.
\newblock {\url{https://doi.org/10.1016/S0370-2693(03)00748-2}}.

\bibitem[Huang and Shovkovy(2003)]{Huang:2003xd}
Huang, M.; Shovkovy, I.
\newblock {Gapless color superconductivity at zero and at finite temperature}.
\newblock {\em Nucl. Phys. A} {\bf 2003}, {\em 729},~835--863.
\newblock {\url{https://doi.org/10.1016/j.nuclphysa.2003.10.005}}.

\end{thebibliography}


\PublishersNote{}
\end{adjustwidth}
\end{document}